\documentclass{emulateapj} 
\usepackage{graphicx}
\usepackage{color,calc}
\usepackage{amsmath}
\usepackage{amssymb}
\usepackage{amstext}

\newcommand{\Fewbody}{{\em Fewbody\/}}

\begin{document}


\title{On the Possibility of Tidal Formation of Binary Planets Around 
Ordinary Stars}
\shorttitle{The Formation of Binary Planets}
\submitted{Submitted to ApJ}
\author{Philipp Podsiadlowski\altaffilmark{1}, Saul Rappaport\altaffilmark{2}, 
John M. Fregeau\altaffilmark{3,4}, Rosemary A. Mardling\altaffilmark{5}}
\shortauthors{Podsiadlowski, et al.}
\altaffiltext{1}{University of Oxford, Department of Astrophysics, Keble Road,
Oxford, OX1 3RH, UK; podsi@astro.ox.ac.uk}
\altaffiltext{2}{M.I.T., Department of Physics and Kavli Institute
for Astrophysics and Space Research, 70 Vassar St., Cambridge, MA
02139, USA; sar@mit.edu}
\altaffiltext{3}{Kavli Institute for Theoretical Physics, UCSB, 
Santa Barbara, CA 93106, USA; fregeau@kitp.ucsb.edu}
\altaffiltext{4}{Chandra/Einstein Fellow}
\altaffiltext{5}{School of Mathematical Sciences, Monash University, 
Clayton, Victoria 3800, Australia; rosemary.mardling@sci.monash.edu.au}

\begin{abstract}
The planet formation process and subsequent planet migration may lead
to configurations resulting in strong dynamical interactions among the
various planets. Well-studied possible outcomes include collisions
between planets, scattering events that eject one or more of the
planets, and a collision of one or more of the planets with the parent
star.  In this work we consider one other possibility that has
seemingly been overlooked in the various scattering calculations
presented in the literature: the tidal capture of two planets which
leads to the formation of a binary planet (or binary brown dwarf) in
orbit about the parent star.  We carry out extensive numerical
simulations of such dynamical and tidal interactions to explore the
parameter space for the formation of such binary planets.  We show
that tidal formation of binary planets is possible for typical planet
masses and distances from the host star.  The detection (or lack
thereof) of planet--planet binaries can thus be used to constrain the
properties of planetary systems, including their mutual spacing during
formation, and the fraction of close planets in very eccentric orbits
which are believed to form by a closely related process.

\end{abstract}



\section{Introduction}

The Earth--Moon and the Charon--Pluto systems are sometimes referred
to as double (or binary) planets, i.e., binary systems consisting of
two planets whose center of mass orbits a central star. These rocky
systems are likely to have formed by the fissioning of a more massive
planet due to a giant impact
\citep{1975Icar...24..504H,Lin1981,2001Natur.412..708C}.  In this
study, we are interested in binaries of gas giant planets (or even
brown dwarfs) for which a fission origin is unlikely. Specifically, we
investigate whether such systems can form by tidal interactions and
their implications for planet formation.  The discovery of such
systems in current or future searches for planet transits (such as the
Kepler [\citealt{kepler2005}] and CoRoT missions
[\citealt{corot2009}]) is exciting for a number of reasons.  These
include (i) the fact that the existence of binary giant planets could
provide strong observational evidence for tidal capture as a viable
astrophysical mechanism.  (ii) Binary planets would allow for new and
important tests and models of planetary dynamics early in the
formation process of planetary systems, including their mutual spacing
during formation.  (iii) They would also allow for studies of
long-term planetary dynamics, including current measures of internal
structure via apsidal motion, and spin-orbit interactions.  (iv)
Binary planets would provide unprecedented accuracy for determining
masses, radii, internal structure, etc.  (v) It seems quite possible
that an eclipsing set of planets, transiting a parent star, would
provide more information on the oblateness of the planets than a
simple transit light curve (see, e.g., \citealt{Carter2010}),
especially given that such a system would be expected to be rotating
more rapidly than a single planet orbiting the parent star. (vi)
Finally, if there is any significant amount of ``magnetic braking" in
a close binary planet system, the two planets could actually be driven
into Roche lobe contact, leading to mass transfer between the planets.
Such a planetary mass-transfer system would most likely be quite
stable and very long lasting.

The formation of a binary planet is intimately linked to the evolution
of the protoplanetary disk from which the planets have formed and
within which they evolve. When the two planets are still embedded in a
gaseous disk, they may migrate inwards or outwards, transferring
orbital angular momentum to or from the disk. If the migration is
relatively slow, the two planets may evolve into an isolated low-order
mean-motion resonance and will then migrate together locked in this
resonance \citep{2002ApJ...567..596L,Papaloizou2005}.  Such a stable
configuration precludes the formation of binary.  However, if
sufficient eccentricity is somehow induced during the migration
process, or if the migration process is sufficiently fast to push the
system through the resonance, neighboring resonances can destabilize
the system.  More generally, the orbits of two planets become
dynamically unstable when the fractional difference of their orbital
radii becomes sufficiently small \citep{1993Icar..106..247G}. The
presence of a disk can inhibit the development of an instability by
limiting the growth of eccentrities.  But once the stabilizing
influence of the disk disappears, e.g., because of a decrease of the
disk mass (more specifically, the disk surface density) or because the
planets have grown sufficiently by accretion from the disk, the planet
orbits can become unstable. This will generally lead to a dynamical
scattering event, in which the configuration of the planets can change
drastically. \citet{2008ApJ...686..621F} and
\citet{2008ApJ...686..580C} have systematically studied such
scattering events in systems with two and three planets and found that
the results of such scatterings could be: (a) the collision and merger
of the two planets, (b) the collision and merger of a planet and the
host star, (c) the ejection of one of the planets, or (d) a
quasi-stable configuration in which both planets remain orbiting the
host star after $\sim 10^6$ orbits. However, these authors did not
consider another possibility: the formation of a binary planet. The
latter may result either from a three-body exchange or more simply
from a tidal capture \citep{1975MNRAS.172P..15F}.

A tidal capture occurs when the two planets get sufficiently close --
typically within a few planet radii -- but do not collide
directly. Such close encounters induce tidal oscillations in one or
both planets, converting orbital energy into oscillation energy (which
is eventually dissipated as thermal energy) and leaving a bound binary
planet (at least temporarily) in orbit about the host star. During
subsequent periastron passages, energy may be exchanged either to or
from the tides (unless the tidal oscillations have been completely
damped in the mean time), and the evolution is formally chaotic
\citep{Mardling1995} in much the same way as an unstable three-body
orbit (again with the possibility of dissociation).  During this
phase, the tides can be extremely large (because the tidal amplitude
is additive), and it is likely that non-linear fluid processes such as
shocks operate to make energy dissipation quite efficient, limiting
the amplitude of the tides.  However, once the system has dissipated a
sufficient amount of tidal energy, it will cease to behave chaotically
and will circularize in the normal fashion with the tidal amplitude
never being able to grow \citep{Mardling1995}.  In the case that tidal
capture occurs in the tidal field of a third body (in the present
case, the star), the tidal energy exchange process can be {\it doubly}
stochastic. However, to simplify the modeling process, we will assume
in this work that all energy deposited in the tides is dissipated
before the next periastron passage, consistent with severe tidal
damping during this phase as discussed above.  The validity of this
assumption depends on the efficiency of the tidal dissipation process,
often parametrized using the tidal $Q$ value (e.g.,
\citealt{Goldreich1966}).

In a three-body exchange, a binary planet forms by the ``exchange'' of
the outer planet into the ``inner binary'' (the central star + inner
planet) forming a new inner binary (the two planets).  If there is no
dissipation of orbital energy, the newly formed pair is susceptible to
dissociation because, unlike a normal two-body tidal capture, it is
continuously forced by the tidal field of the star. The binary orbit
will exchange energy and angular momentum with its center-of-mass
orbit around the star in a random-walk fashion (the system is formally
chaotic; \citealt{2008LNP...760...59M}), until sufficient energy has
been transferred to the binary to dissociate it.  This can only be
avoided once enough orbital energy is {\it permanently} removed from
the binary to make the system stable against chaotic interactions.
This can occur through (point-mass) interactions with other bodies
(e.g., a third planet or a background of planetesimals), similar to
the mechanism proposed for the formation of Kuiper-Belt binaries
\citep{2002Natur.420..643G}, or through tidal dissipation if the
planets are close enough.

In both scenarios, tidal dissipation may play a key role in forming a
binary planet, but, in the first case, the tidal capture has to operate
in one (or possibly a few) encounters, while, in the second, tidal
dissipation may operate over the much longer timescale of the transient
binary state. We therefore expect that, in the first case, the
post-capture orbital separation of the binary planet is a few planet
radii, while, in the second, it could be as wide as is stable on a long
time timescale.

In this study, we consider primarily the mutual tidal capture of two
planets in a ``dynamically active'' planetary system. We show that,
for reasonable assumptions about the tidal coupling and dissipation of
gas giant planets, mutual tidal capture of planets into a
planet--planet binary is a relatively generic feature of dynamically
systems.  In this regard, it has recently been shown that the
long-standing issue of the relatively high eccentricities of
extrasolar planets can be understood if the configuration of the
newly-formed planetary system after the gas disk has dissipated is
dynamically active, so that planet--planet scattering can operate
and eject planets while increasing the eccentricities of the remaining
planets
\citep{2008ApJ...686..621F,2008ApJ...686..603J,2008ApJ...686..580C}.
It is not obvious that planetary systems forming via the usual core
accretion scenario should result in dynamically active systems after
the gas disk phase.  However, simulations starting with protoplanets
in a gas disk and following both gas physics and $N$-body dynamics
have shown that such resulting configurations are indeed possible
\citep{2008Sci...321..814T,2010ApJ...714..194M}.  

Based on our calculations, we present the {\em relative} probabilities
of the possible outcomes of dynamically active systems (including,
e.g., planet--planet collisions, tidal capture, ejection).  We discuss
the potential observability of a planet binary in relation to the
likelihood that planetary systems are dynamically active early in
their lifetimes.

In our study we also allow for the possibility of tidal capture of
{\em brown dwarfs} (BDs) in close orbit around a hydrogen-burning
star.  Brown dwarfs probably form in a very different way from, and at a
different distance than, gas giant planets.  While gas giants likely
form via core accretion, any brown dwarf forming out of a
circumstellar disk will do so via gravitational fragmentation, and at
fairly large distances from the host star ($\gtrsim 10^2 \, {\rm AU}$),
where the disk is Toomre unstable \citep{Toomre1964} and can cool
sufficiently so that unstable clumps can collapse (see, e.g.,
\citealt{2009MNRAS.392..413S}; also see \citealt{Boss1997}).  We do not
address the issue of how to get the brown dwarfs to distances of
$\lesssim 1\, {\rm AU}$ from the host star, but note that such systems
are known to exist (e.g., CoRoT 15b contains a brown dwarf orbiting an F
star with an orbital period of 3\,d; \citealt{bouchy2010}). In any
case, as we will show, if multiple brown dwarfs exist on such close
orbits, they can easily become bound in a BD--BD binary via tidal
capture.\footnote{Throughout this paper, we do not sharply distinguish
  between planets and brown dwarfs and often refer to all types of
  sub-stellar objects collectively as planets.}

In \S 2 we present the results of numerical simulations of dynamically
active planet systems for a range of masses and distances from the
parent star.  In \S 3 we describe the detectability of such binary
planets in radial velocity searches and in transit studies.

\section{Numerical Simulations}\label{sec:simulations}

\subsection{Numerical Method}

Since the dynamical processes leading to planet--planet tidal capture
are the same as those that lead to planet--planet collisions, any
numerical method which can properly treat ``dynamically active''
planetary systems should serve as an appropriate base upon which we
can add a treatment of tidal dissipation.  We use the
\Fewbody\ integrator, which is designed for strong small-$N$-body
gravitational encounters \citep{2004MNRAS.352....1F}. We have
tested that the code is suitable for simulating dynamically active
planetary systems by comparing our results to the simulations of 
\citet{2008ApJ...686..621F} (see Appendix).

In order to include tidal capture, we also had to include a treatment
of tidal dissipation in \Fewbody.  We use the functional fits to the
energy dissipated by the close passage of two polytropes as presented
in \citet{1993A&A...280..174P}. When the two planets, of mass $m_i$
and radius $R_i$, encounter each other with close approach distance
less than $10(R_1+R_2)$, we calculate the energy dissipated in the
encounter by treating each planet as an $n=3/2$ polytrope and reduce
the planet--planet relative velocity (at pericenter) accordingly.  We
only include tidal dissipation in impulsive interactions, requiring --
somewhat arbitrarily -- that the close approach distance is less than
10\% of the previous local maximum in the planet--planet
distance.\footnote{This means that no tidal damping is applied for
  eccentricity $\la 0.8$ in a bound planet-planet system.}  This
provides a lower limit on the overall tidal dissipation.  If the close
approach distance is less than $R_1+R_2$ we assume the planets merge
instantaneously, conserving mass and momentum.

\vspace{0.3cm}

\subsection{Stability Criterion}

\Fewbody\ automatically terminates a calculation when an unambiguous,
dynamically stable configuration has been obtained.  To accurately
test for the dynamical stability of a binary planet orbiting a star,
we use the newly improved algorithm of \citet{2008LNP...760...59M},
which utilizes the concept of orbital resonance overlap to test for chaos
and, ultimately, instability.  The resonance overlap algorithm in its
current form is only approximate for small values of the outer
eccentricity $e_{\rm out}$, i.e., the eccentricity of the outermost
body (in this context the parent star) about the center of
mass of the innermost binary (the planet-planet binary).  We thus set
$e_{\rm out}=0.1$ in the algorithm when the outer eccentricity is
smaller than this value.  In addition, we also require that the
apocenter of the planet--planet orbit is within 2/3 of its Hill radius
at pericenter.  Specifically, we require
\begin{equation}
  a_{\rm in} (1+e_{\rm in}) < \frac{2}{3} a_{\rm out}(1-e_{\rm out}) 
\left(\frac{m_1+m_2}{3M_\star}\right)^{1/3} \, ,
\end{equation}
where $a_{\rm in}$ and $e_{\rm in}$ are the semi-major axis and
eccentricity of the newly formed planet--planet binary, and the
expression on the right is 2/3 the radius of the Hill sphere of the
planet--planet pair evaluated at apocenter.

\subsection{Starting Conditions}

We adopt initial conditions for our simulations that are similar to
those of \citet{2008ApJ...686..621F}.  The planet masses are either
$m_1=10^{-3}\, M_\sun$ and $m_2=0.5 \times 10^{-3}\, M_\sun$ for the
``gas giant'' case, or $m_1=70 \times 10^{-3}\, M_\sun$ and $m_2=30
\times 10^{-3}\, M_\sun$ for the ``brown dwarf'' case.  In all cases,
we fix the central star's mass at $M_\star = 1 \, M_\sun$, and planet
1 initially orbits with semi-major axis $a_{\rm 1,init} < a_{\rm
  2,init}$.  We use three initial values for $a_1$: $a_{\rm
  1,init}=0.2$, $1$, and $5\, {\rm AU}$.  The initial semi-major axis
of planet 2 is set so that the system is not Hill stable, with $a_{\rm
  2,init}$ randomly distributed uniformly between $0.9a_{\rm
  1,init}(1+\Delta_c)$ and $a_{\rm 1,init}(1+\Delta_c)$, where
$\Delta_c = 2.4 (m_1/M_\star+m_2/M_\star)^{1/3}$
\citep{1993Icar..106..247G}.  The planet initial eccentricities
$e_{\rm i,init}$ are randomly distributed uniformly between 0 and
0.05, and the relative inclination of the planet orbits is distributed
uniformly between $0^\circ$ and $2^\circ$.  All remaining orbital
elements are sampled uniformly in their allowed range.  For reference,
all model parameters are shown in Table \ref{tab:models}.  

\begin{figure}
  \begin{center}
    \includegraphics[width=0.95\columnwidth]{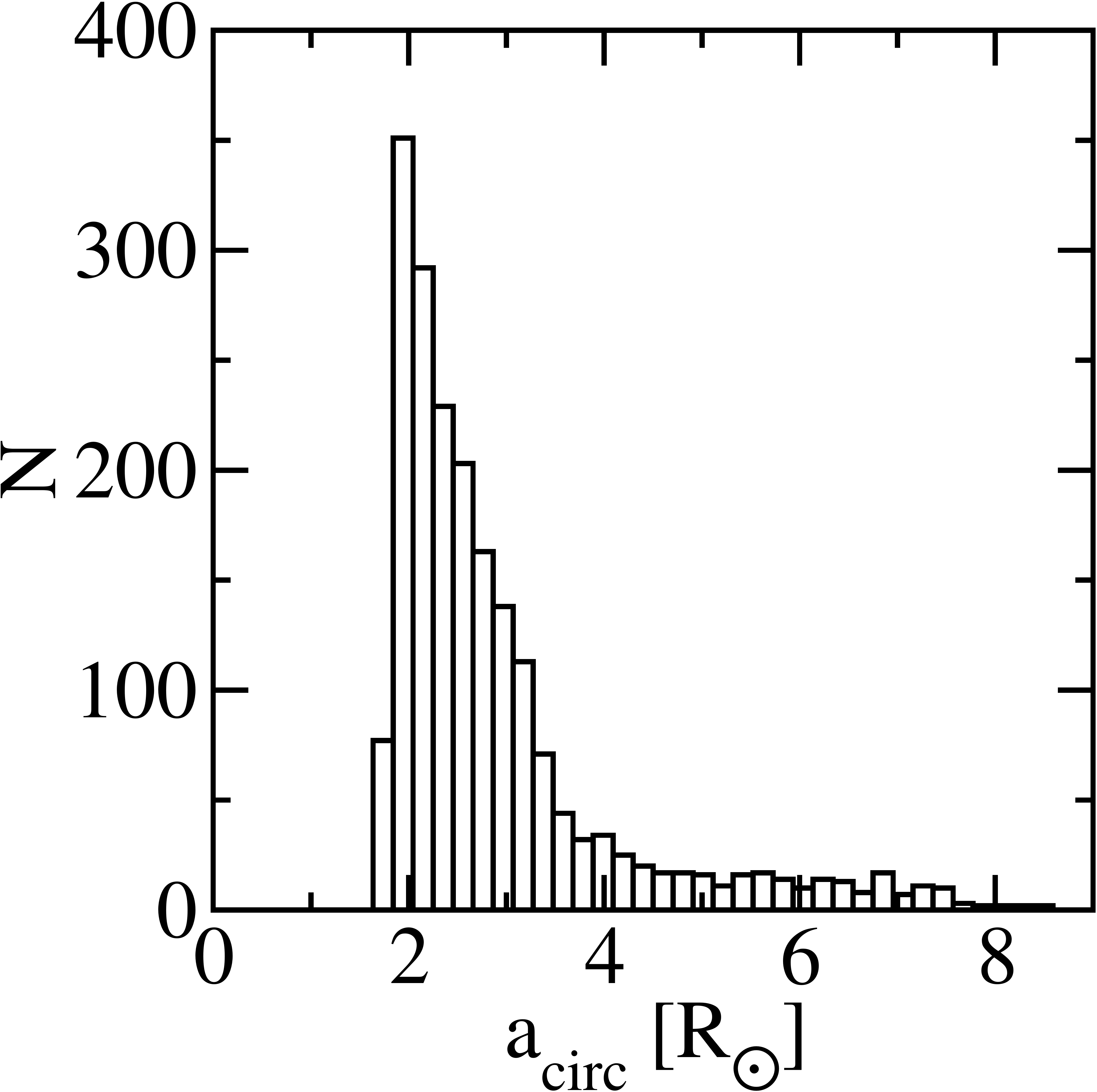}
    \caption{Histogram of semi-major axes of circularized binary brown dwarfs
resulting from tidal capture events in model t21 (with
brown dwarfs masses of 30\,$M_{\rm J}$ and 70\,$M_{\rm J}$, respectively).
      \label{fig:hist}}
  \end{center}
\end{figure}

\begin{figure}
  \begin{center}
    \includegraphics[width=0.95\columnwidth]{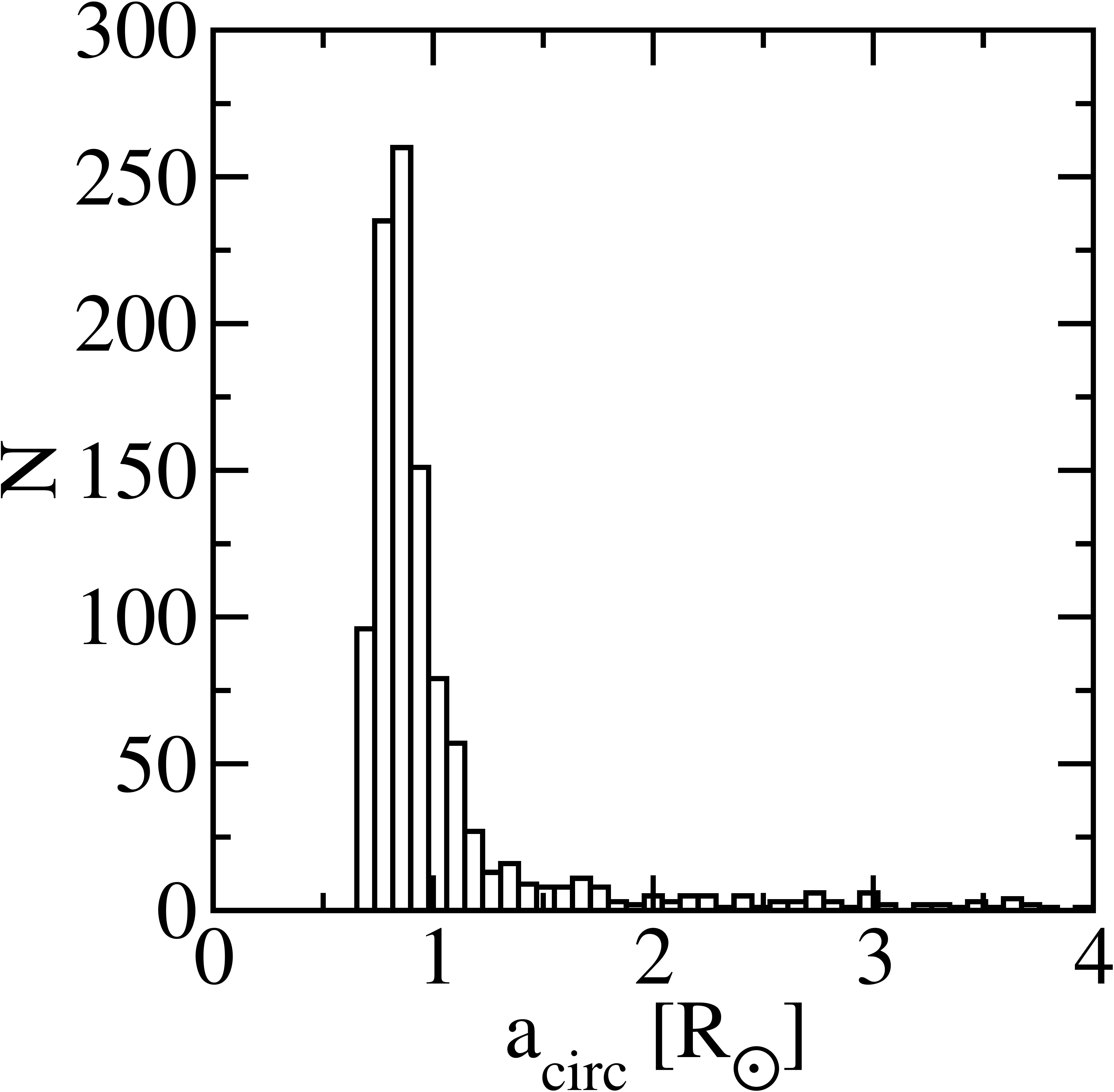}
    \caption{Histogram of semi-major axes of circularized binary
planets resulting from tidal capture events in model t16 (with
planet masses of 1\,$M_{\rm J}$ and 0.5\,$M_{\rm J}$, respectively).
      \label{fig:hist2}}
  \end{center}
\end{figure}

\subsection{Thermal Bloating of the Interacting Planets}

\begin{figure}
  \begin{center}
    \includegraphics[width=0.95\columnwidth]{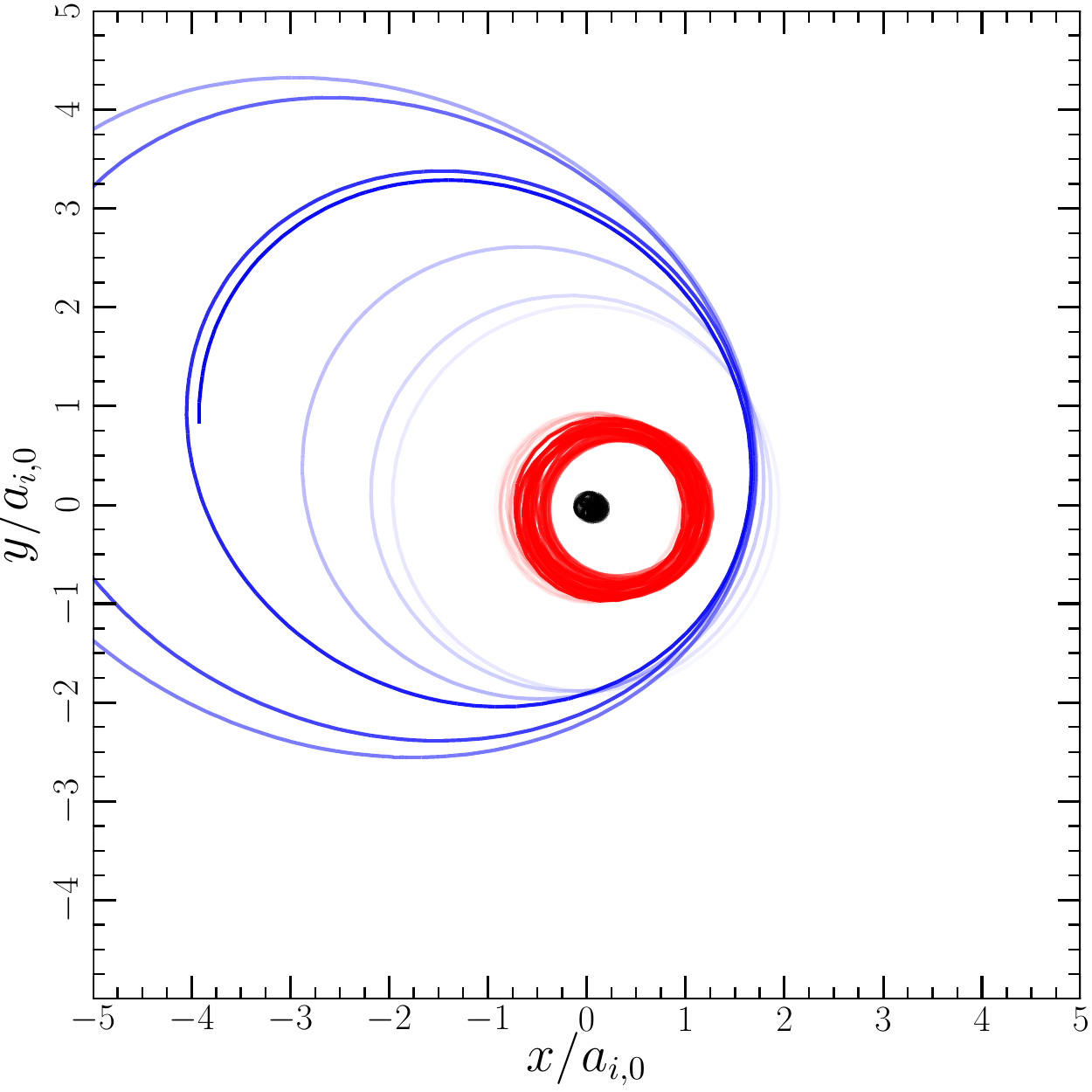}\vglue1pt
    \includegraphics[width=0.95\columnwidth]{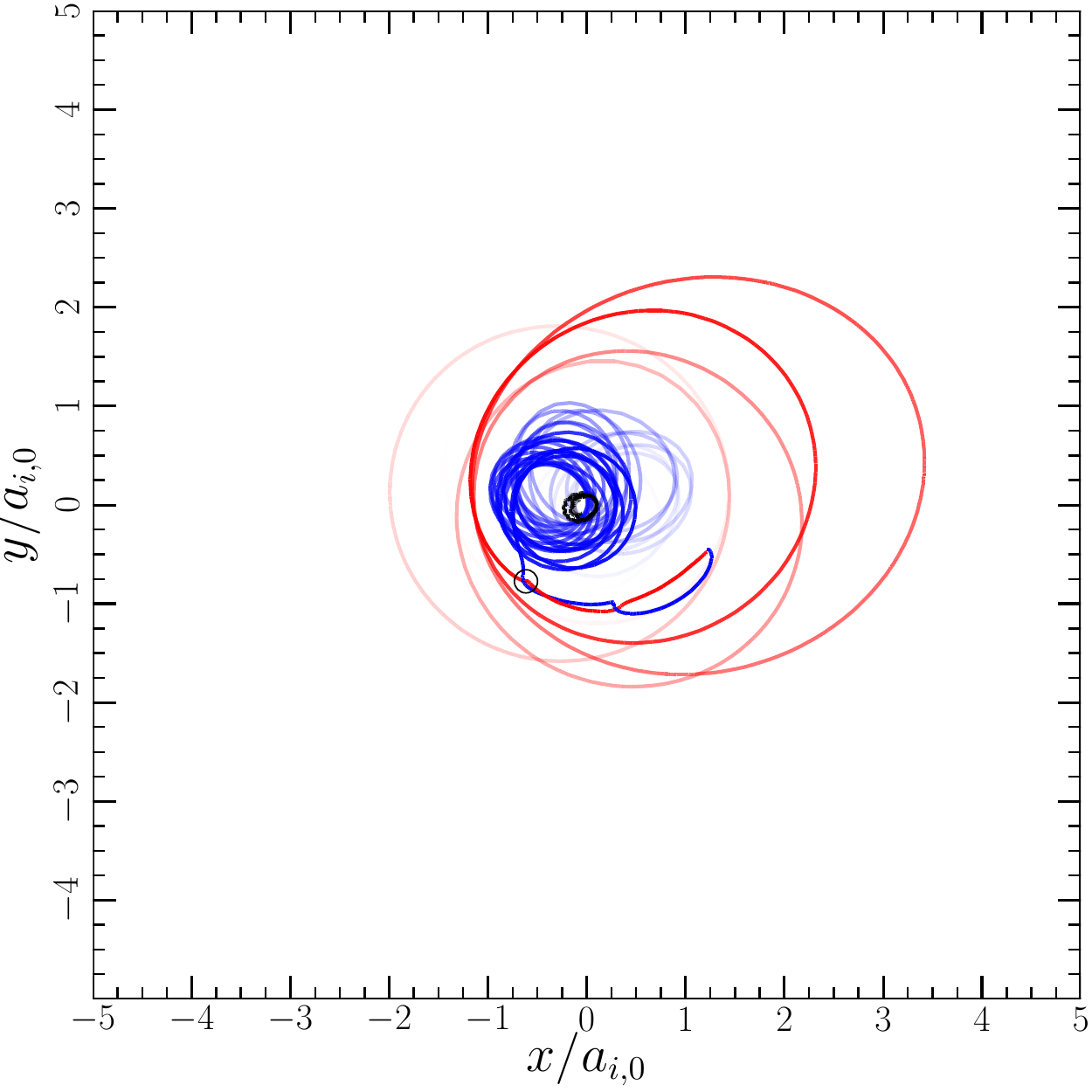}
    \caption{Evolution of a planet/brown dwarf binary system from model t21
      in the $x$--$y$ plane at the start of the calculation (upper
      panel), and at the end of the calculation (lower panel).  The
      star is represented by a black trail, and the planets by red and
      blue trails.  The trails fade with time so that the darkest
      points along a curve are the most recent.  Dissipative tidal
      encounters are shown as open circles at the point where they
      occur.  The system becomes active quickly after the start of the
      calculation.  Over time the planets exchange position relative
      to the star.  The planets suffer a weak dissipative tidal
      encounter at late time but eventually collide and merge.
      \label{fig:bdmerger}}
  \end{center}
\end{figure}

\begin{figure}[h]
  \begin{center}
    \includegraphics[width=0.95\columnwidth]{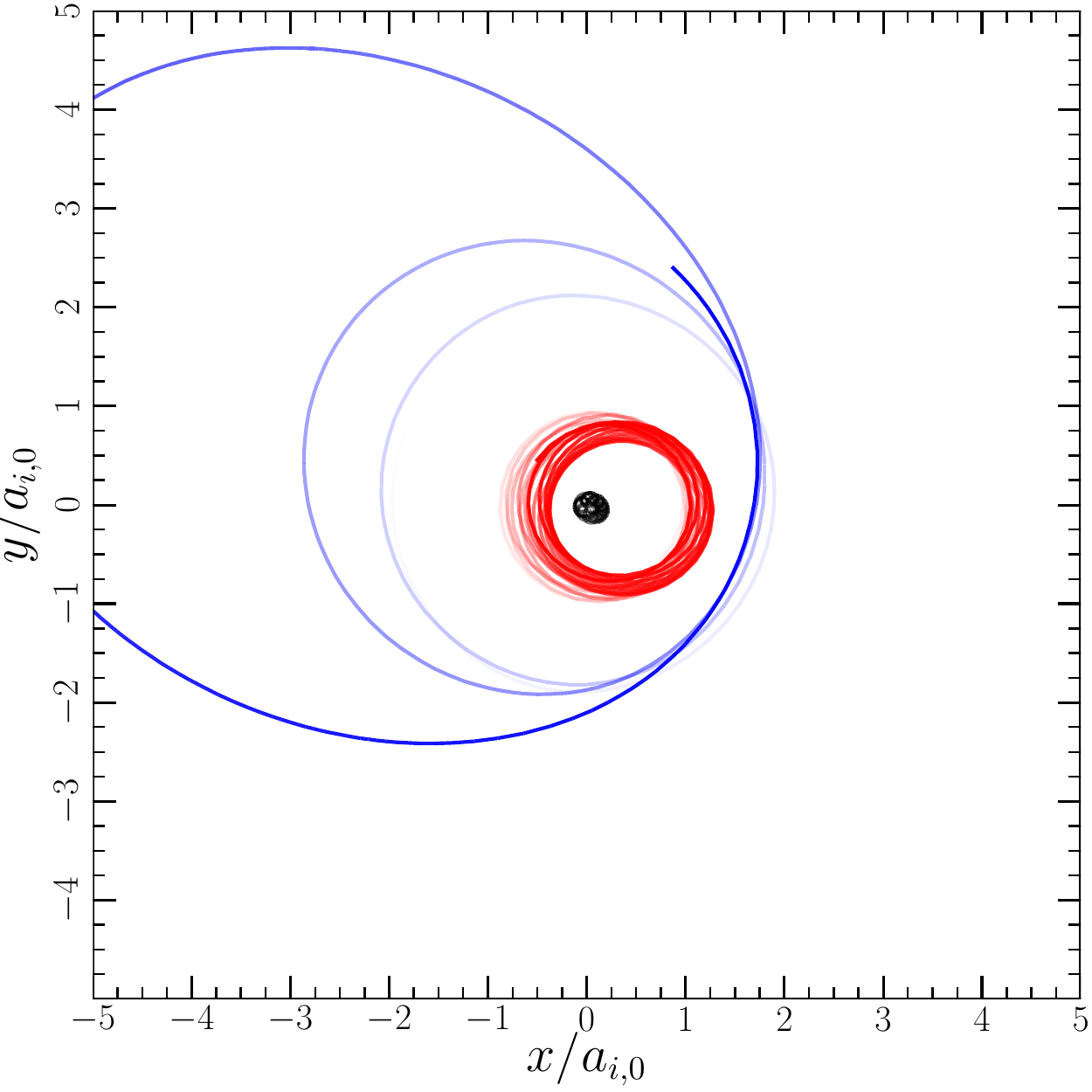}\vglue1pt
    \includegraphics[width=0.95\columnwidth]{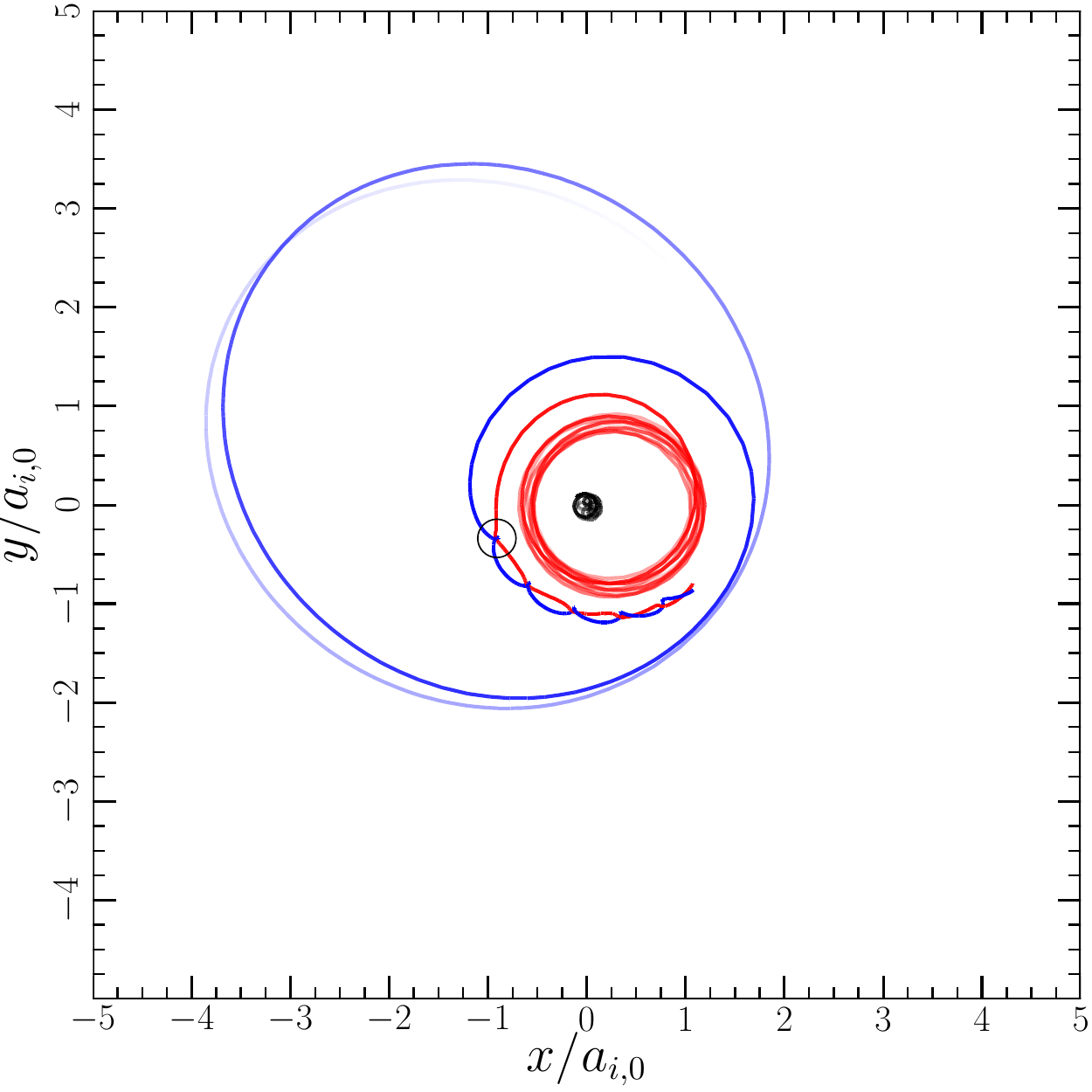}
    \caption{Evolution of a planet/brown dwarf binary system from model t21
      resulting in tidal capture.  Conventions are as in
      Fig.\ \ref{fig:bdmerger}.  The system becomes dynamically active
      shortly after the calculation begins.  A strong tidal encounter
      at a later time binds the planet--planet pair, resulting in a
      configuration that is dynamically stable.  The planet--planet
      binary's circularized semi-major axis is $2.3\, R_\sun$.
      \label{fig:bdtc}}
  \end{center}
\end{figure}

In this study, we assume that each of the pair of planets/brown dwarfs
was born early in the evolutionary history of the parent star.  In
particular, we adopt the hypothesis that the planets/brown dwarfs were
either born separately in the unstable collapse of a massive accretion
disk, or otherwise were driven to migration toward the parent star via
an accretion disk.  In either case, the planets or brown dwarfs were
necessarily young when they captured each other, i.e., had an age of
$\sim 1-30$ Myr.  Planets, and especially brown dwarfs, are quite
thermally bloated at these young ages (see, e.g., Fig.\,4 of
\citealt{NRJ93}).  We have constructed a semi-empirical fitting
formula for the radius of brown-dwarfs of mass $\gtrsim 15 \,M_{\rm
  J}$ as a function of evolution time, $t$, as follows:
\begin{equation}
R(t) =0.79 \sqrt{M_{i}/M_{\rm J}}\,t_6^{-1/3} \,R_{\rm J}\,,
\label{eq:bloat}
\end{equation}
where $t_6$ is the evolution time in units of $10^6$\,yr, $M_i$ is the
mass of the planet, and $M_{\rm J}$ and $R_{\rm J}$ are the mass and
radius of Jupiter, respectively.  This expression works quite well for
masses above $\sim 20\,M_{\rm J}$ and for ages in the range of
$0.1$\,--\,$30$\,Myr.  For any object whose radius falls below
$1\,R_{\rm J}$ based on this expression, we simply fix the radius at
$1\,R_{\rm J}$.\footnote{Note that this prescription may underestimate
the radii of very close Jupiter-mass planets that could be inflated
due to tidal heating \citep{Bodenheimer2001}.}

Tidal dissipation is a strong function of the radius of the planet or
brown dwarf compared to its separation from the object with which it
is interacting.  Hence, the inclusion of thermal bloating is
potentially important, as it allows tidal capture to occur at larger
initial separations.

After a planet--planet binary is formed via tidal capture and the
resulting star--planet--planet system is deemed dynamically stable by
the \citet{2008LNP...760...59M} stability criterion, we stop the
calculation and record the semi-major axis, $a_{\rm in}$, and
eccentricity, $e_{\rm in}$, of the planet--planet binary.  In
post-processing we assume the orbit is quickly tidally circularized,
and set $a_{\rm circ}=a_{\rm in} (1-e_{\rm in}^2)$.
Fig.\ \ref{fig:hist} shows a histogram of the circularized
planet--planet semi-major axes resulting from tidal capture events in
model t21 (for a pair of brown dwarfs).  There is a clear peak just
above $a_{\rm circ} \approx 2\, R_\sun$, with a tail that extends out
to $\sim 8\, R_\sun$.  Fig.\ \ref{fig:hist2} shows the distribution
for model t16 (for a pair of giant gas planets).

\subsection{Illustrative Scattering Results}

To get a better feel for how the dynamics unfolds in a simulation
resulting in tidal capture, we have plotted the evolution of the star
and `planet' positions for a set of representative simulations from
model t21.\footnote{Model t21 is a simulation for brown dwarfs, but the
scattering results for the planet simulations are very similar.}
Fig.\ \ref{fig:bdmerger} shows a typical simulation ending
in a {\em merger} of the two planets.  The system becomes active
quickly after the start of the calculation.  Over time the planets
exchange position relative to the star.  The planets suffer a weak
dissipative tidal encounter at late time but eventually collide and
merge.

\begin{figure}[h]
  \begin{center}
    \includegraphics[width=0.95\columnwidth]{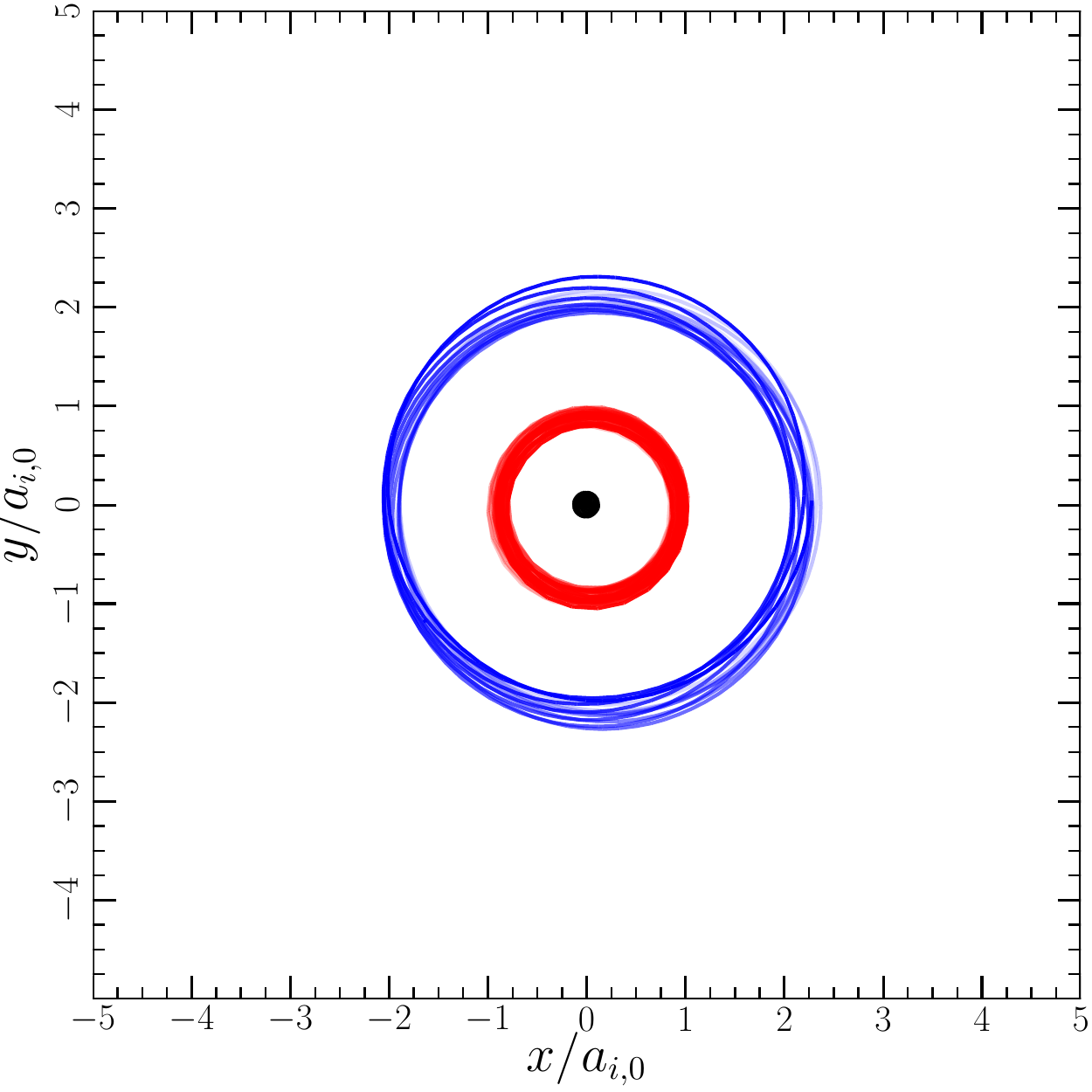}\vglue1pt
    \includegraphics[width=0.95\columnwidth]{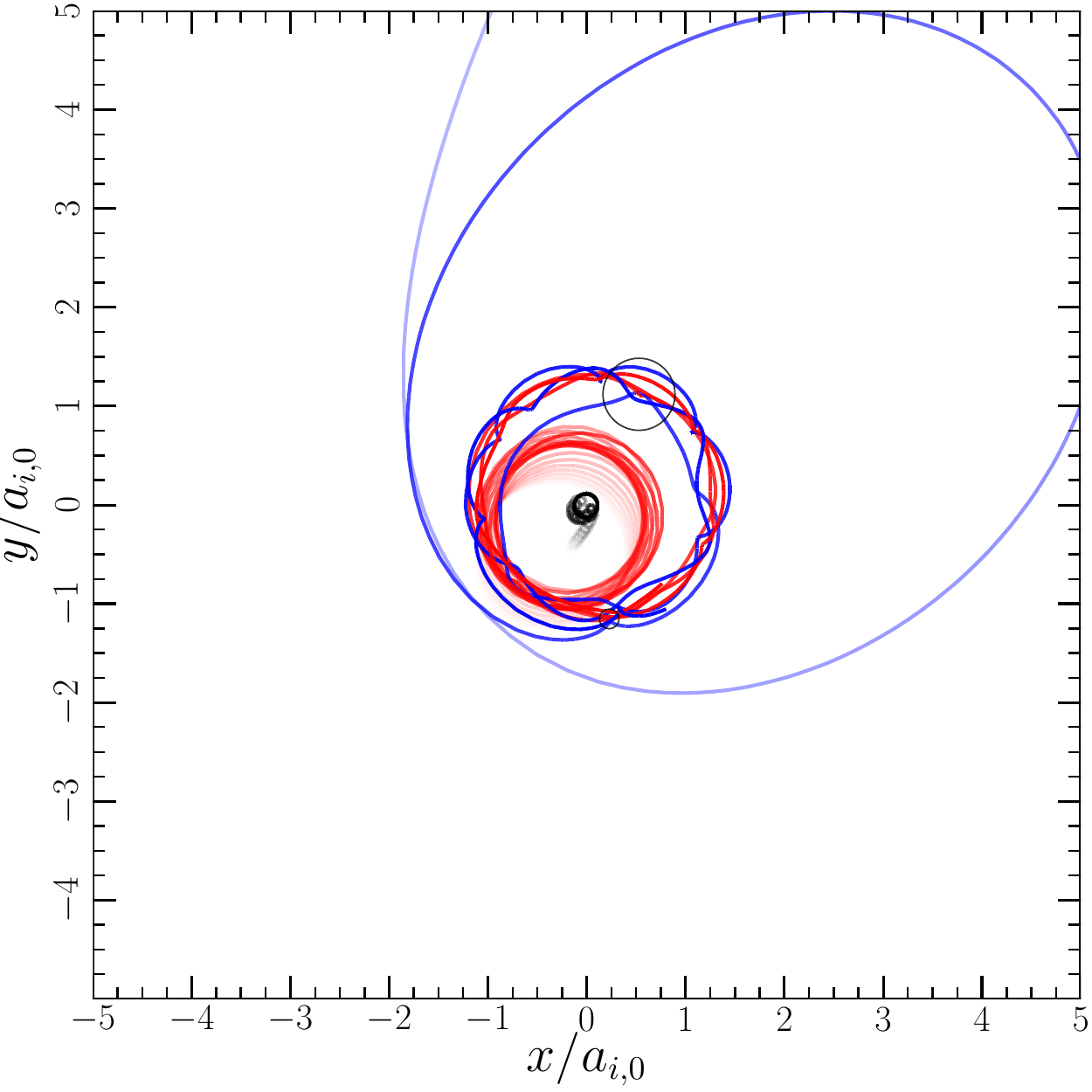}
    \caption{Evolution of a planet/brown dwarf binary system from model t21
      resulting in a somewhat gradual tidal capture.  Conventions are
      as in Fig.\ \ref{fig:bdmerger}.  A weak tidal encounter (the
      smaller of the two open circles) nearly results in a dynamically
      stable planet--planet binary.  The planets later suffer a
      stronger tidal encounter (the larger of the two open circles)
      that results in a stable configuration.  Note that the radius of
      the open circles is proportional to the fractional change in
      relative planet--planet velocity resulting from the encounter.
      In this case, the planet--planet binary's circularized
      semi-major axis is $6.0\, R_\sun$.
      \label{fig:bdtcg}}
  \end{center}
\end{figure}

Fig.\ \ref{fig:bdtc} shows a typical {\em ``direct'' tidal capture}
simulation.  The system becomes dynamically active shortly after the
calculation begins.  A strong tidal encounter at a later time binds
the planet--planet pair, resulting in a configuration that is
dynamically stable.  The planet--planet binary's circularized
semi-major axis is $2.3\, R_\sun$.  Fig.\ \ref{fig:bdtcg}, on the
other hand, shows a typical {\em ``gradual'' tidal capture}
simulation.  A weak tidal encounter (the smaller of the two open
circles) nearly results in a dynamically stable planet--planet binary.
The planets later suffer a stronger tidal encounter (the larger of the
two open circles) that results in a stable configuration.  In this
case, the planet--planet binary's circularized semi-major axis is
$6.0\, R_\sun$. These ``gradual'' tidal captures are responsible for
the long tails in the final semi-major axis distributions in
Figs.~1 and 2.

\subsection{Summary of Scattering Results}

\begin{figure}[h]
  \begin{center}
    \includegraphics[width=0.99\columnwidth]{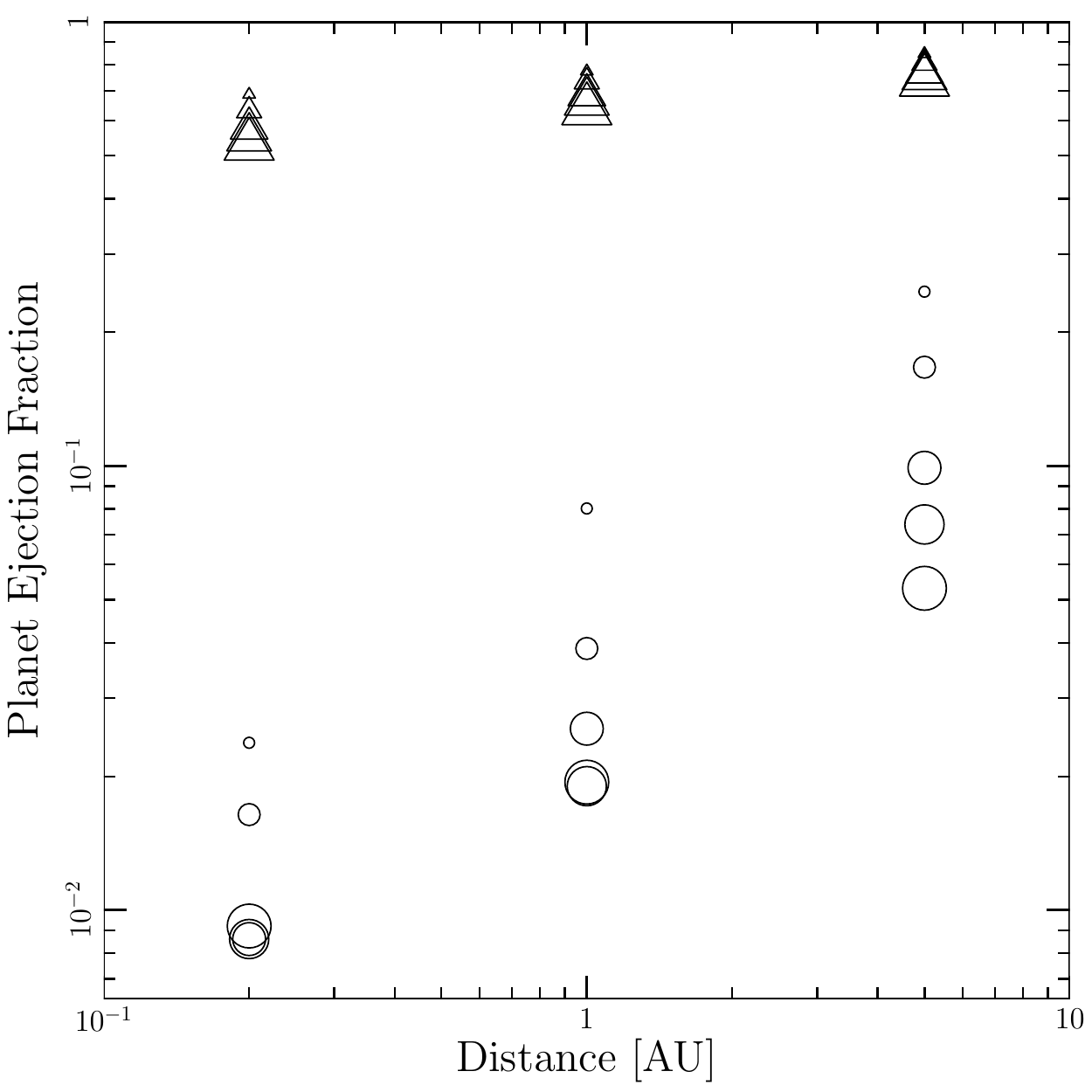}
    \caption{The fraction of dynamically active planet systems in
      which one of the planets is ejected as a function of the
      distance from the parent star.  Triangles are for
      ``planets'' in the brown-dwarf mass range, while circles
      represent gas giants of approximately Jupiter's mass.
      Symbol size is proportional to the log of the planet/brown dwarf
      radius.
      \label{fig:ejec}}
  \end{center}
\end{figure}

\begin{figure}[h]
  \begin{center}
    \includegraphics[width=0.99\columnwidth]{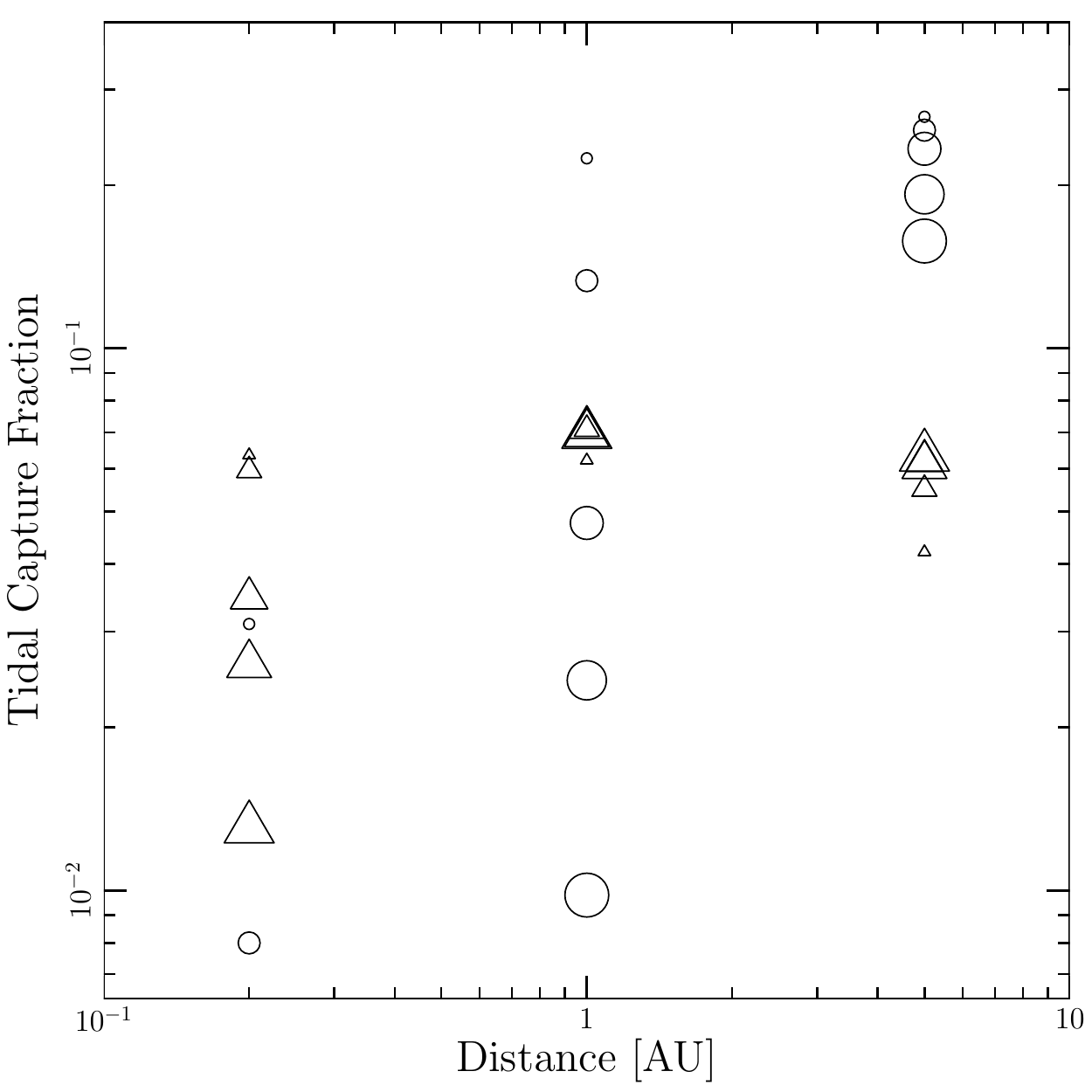}
    \caption{The fraction of dynamically active planet systems in
      which a successful tidal capture occurs leading to the formation
      of a stable binary planet.  Triangles are for ``planets'' in
      the brown-dwarf mass range, while circles represent gas
      giants of approximately Jupiter's mass.  
      Symbol size is proportional to the log of the planet/brown dwarf
      radius.
      \label{fig:tidal}}
  \end{center}
\end{figure}

For reference, we give the outcome statistics for each model in Table
\ref{tab:stats}.  We also show in Fig.\,\ref{fig:ejec} a summary of
the fraction of dynamically active systems in which one of the planets
is ejected as a function of the different assumed distances from the
parent star, and in Fig.\,\ref{fig:tidal} the fraction of systems
leading to a successful tidal capture.

\begin{figure}[h]
  \begin{center}
    \includegraphics[width=0.99\columnwidth]{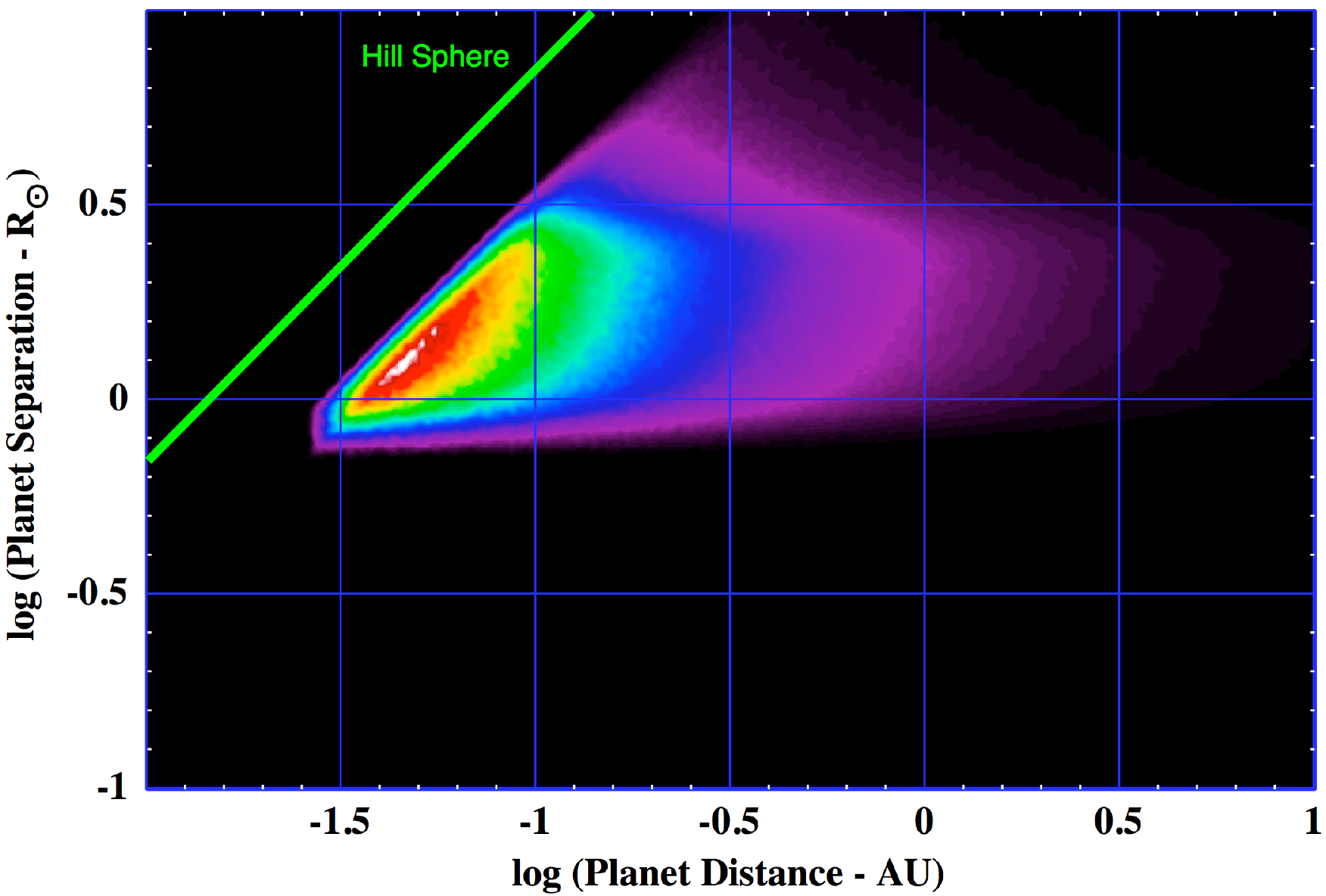}\vglue4pt
    \includegraphics[width=0.99\columnwidth]{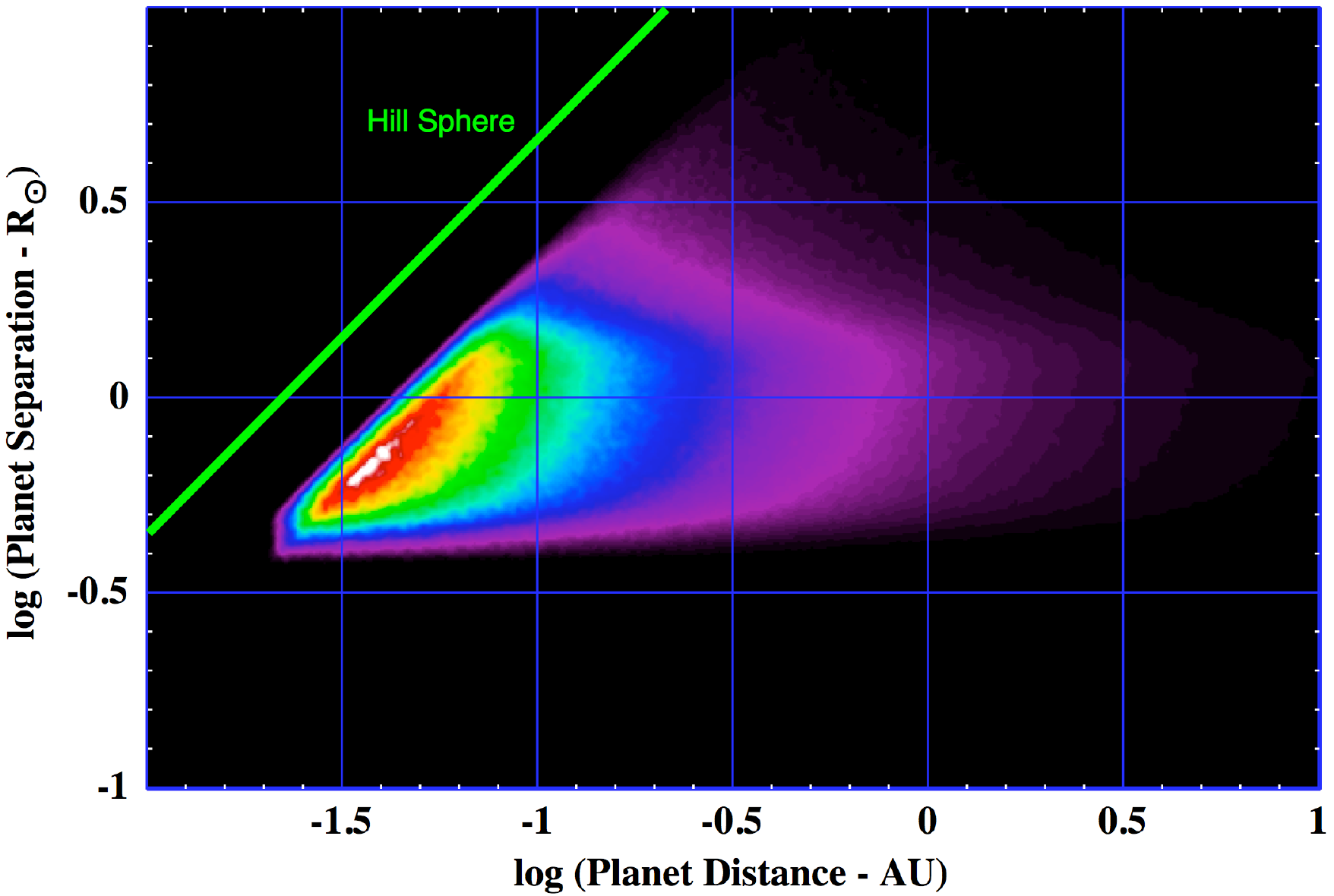}\vglue4pt
    \includegraphics[width=0.99\columnwidth]{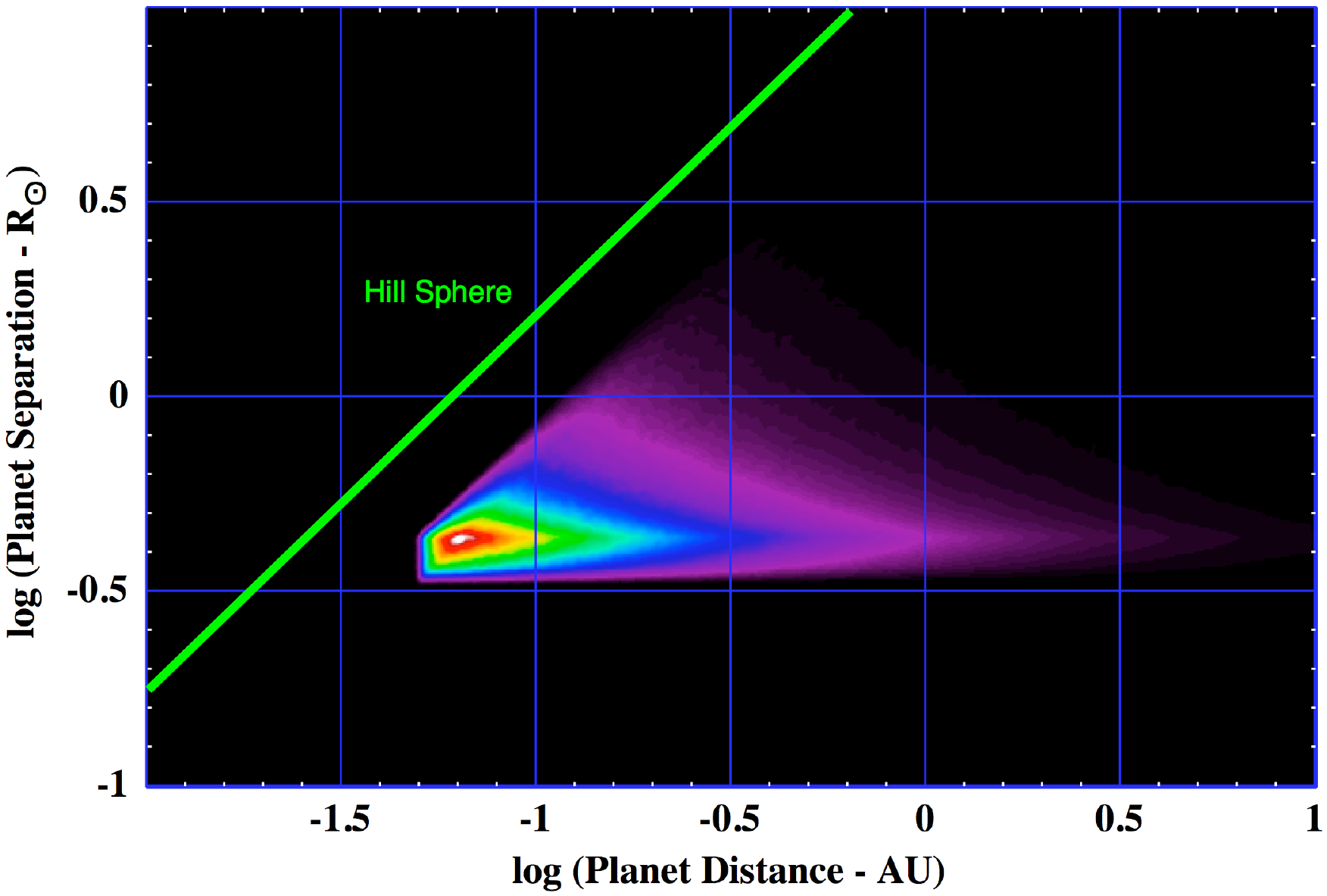}
    \caption{Simulated populations of Jovian binary planets (bottom
      panel), super-Jupiter binary planets (middle panel) and binary
      brown dwarfs (top panel) in orbit about the parent star (in this
      case, $1\,M_\odot$).  The masses of each binary pair are
      $\{70,30\},\{20,10\},\{1,\frac{1}{2}\}$ Jupiter masses (top to
      bottom panels).  The color shading indicates the relative
      probability of discovering such a system, assuming that the
      binaries are born uniformly distributed in $\log d$.  White to
      purple colors indicate probability ratios of $\sim 50:1$. See
      text for specific algorithms used here. The radius of the Hill
      sphere is indicated by the diagonal green line.
      \label{fig:sim}}
  \end{center}
\end{figure}

\subsection{Simulated Population of Binary Planets}

Based on the results of the numerous scattering events that we have
computed for a wide range of masses and distances from the parent
star, we have devised a simple fitting formula that approximates the
distribution for forming a tidally-captured binary pair
of planets or brown dwarfs with a (final) circularized orbital
separation, $a_{\rm circ}$.  Expressed in Monte Carlo form, we have:
\begin{equation}\label{eq:sep}
a_{\rm circ}({X}) = \left(\frac{R_{i\rm, bloat}}{R_{\rm J}}\right)
\left(\frac{1}{3}+\frac{0.1{X}}{1-{X}}\right)~R_\odot\,,
\end{equation}
where $R_{i\rm, bloat}$ is the thermally bloated radius of the larger
of the planets/brown dwarfs at the time of tidal capture, $R_{\rm J}$
is the radius of Jupiter, and $X$ is a uniformly distributed random
number between 0 and 0.99 (the upper limit in $X$ introduces an upper
cutoff of $a_{\rm circ}\simeq 10\,R_\odot$, above which our scattering
experiments provide insufficient statistics).  While the overall
probabilities for tidal capture do vary systematically with distance
from the parent star (see Table \ref{tab:stats}) and the masses of the
scattering objects, the basic functional form of the distribution of
$a_{\rm circ}$, and its dependence on the radii of the tidal capturing
planets, seem relatively independent of distance and masses.

Once we have in place analytic expressions for the thermal bloating as
a function of age, and the distribution of tidal-capture circularized
orbital radii as a function of the thermal bloating, we can use these
to generate a synthetic population of binary planets and brown dwarfs.
In this simulation, we first choose a random location, $d$, for the
planet/brown dwarf binary between 0.01 and 10 AU, uniformly
distributed in $\log d$.  This is completely arbitrary since we do not
know {\em a priori} the initial distribution of where planets and
brown dwarfs form in a protoplanetary disk.  Next, we choose a random
age for the planet/brown dwarf pair between 1 and 15 Myr, uniformly
distributed in $\log t$.  The age sets the size of the thermally
bloated radius according to equation (\ref{eq:bloat}).  The final
(late-age) circularized orbital separation, $a_{\rm circ}$, of the
binary is chosen randomly from the distribution given by equation
(\ref{eq:sep}).  Finally, we choose the eccentricity of the outer
orbit (i.e., the orbit of the CM of the planet/brown-dwarf binary
around the parent star) from the following probability distribution
derived from our scattering studies:
\begin{equation}
p(e) \propto \exp(-e/0.05)\,.
\end{equation}

We require for each binary pair, chosen according to the above
prescriptions, that its orbital separation $a_{\rm circ}$ be smaller
than 40\% of the radius of the Hill sphere for that particular binary
(this depends on the masses and the eccentricity of the outer binary)
in order to ensure long-term stability
\citep{2006MNRAS.373.1227D}.\footnote{Their stability criteria apply
  to the case of massless satellites of planets.  We have performed
  some numerical tests for equal-mass planet--planet binaries in
  circular orbits and obtained similar stability boundaries.}  Once
the parameters of a binary have been fully chosen, we assign a
probability of detection via transits simply as $\propto R_*/d$, where
$d$ is the distance of the binary CM from the parent star of radius
$R_*$.  Obviously, this probability could be enhanced if the projected
separation of the planet/brown-dwarf binary, as seen from the Earth,
is larger than the parent star.  However, such calculations are beyond
the scope of the paper.

The results of our simulations are shown in Fig.\,\ref{fig:sim} for
three different mass pairs of planets/brown dwarfs.  The top, middle,
and bottom panels are for the mass pairs:
$\{70,30\},\{20,10\},\{1,\frac{1}{2}\}$ Jupiter masses, respectively.
In each panel the color shading is proportional to the relative
probability of finding a transiting binary pair at planet distance,
$d$, and planet separation, $a_{\rm circ}$.  The radius of the Hill
sphere is shown as the diagonal green line.

\vspace{0.2cm}

\begin{deluxetable*}{ccccccc}
  \tablecaption{Parameters for planet evolution simulations.\label{tab:models}}
  \tablehead{
    \colhead{model name} &
    \colhead{$m_1/10^{-3}\, M_\sun$} &
    \colhead{$m_2/10^{-3}\, M_\sun$} &
    \colhead{$R_1/R_\sun$} &
    \colhead{$R_2/R_\sun$} &
    \colhead{$a_{\rm 1,init}/{\rm AU}$} &
    \colhead{tidal dissipation}
  }
  \startdata
  t2 & 1 & 1 & 0.1 & 0.1 & 5 & off\\
  t3 & 3 & 3 & 0.1 & 0.1 & 5 & off\\
  t4 & 4 & 2 & 0.1 & 0.1 & 5 & off\\
  t5 & 4 & 2 & 0.1 & 0.1 & 5 & on\\
  t6 & 50 & 25 & 0.1 & 0.1 & 5 & on\\
  t7 & 4 & 2 & 0.1 & 0.1 & 0.2 & on\\
  t8 & 50 & 25 & 0.1 & 0.1 & 0.2 & on\\
  t9 & 1 & 0.5 & 0.1 & 0.1 & 5 & on\\
  t10 & 70 & 30 & 0.1 & 0.1 & 5 & on\\
  t11 & 1 & 0.5 & 0.1 & 0.1 & 1 & on\\
  t12 & 70 & 30 & 0.1 & 0.1 & 1 & on\\
  t13 & 1 & 0.5 & 0.1 & 0.1 & 0.2 & on\\
  t14 & 70 & 30 & 0.1 & 0.1 & 0.2 & on\\
  t15 & 70 & 30 & 0.2 & 0.2 & 0.2 & on\\
  t16 & 1 & 0.5 & 0.2 & 0.2 & 5 & on\\
  t17 & 70 & 30 & 0.2 & 0.2 & 5 & on\\
  t18 & 1 & 0.5 & 0.2 & 0.2 & 1 & on\\
  t19 & 70 & 30 & 0.2 & 0.2 & 1 & on\\
  t20 & 1 & 0.5 & 0.2 & 0.2 & 0.2 & on\\
  t21 & 70 & 30 & 0.6 & 0.4 & 0.2 & on\\
  t22 & 1 & 0.5 & 0.4 & 0.4 & 5 & on\\
  t23 & 70 & 30 & 0.4 & 0.4 & 5 & on\\
  t24 & 1 & 0.5 & 0.4 & 0.4 & 1 & on\\
  t25 & 70 & 30 & 0.4 & 0.4 & 1 & on\\
  t26 & 1 & 0.5 & 0.4 & 0.4 & 0.2 & on\\
  t27 & 70 & 30 & 0.4 & 0.4 & 0.2 & on\\
  t28 & 1 & 0.5 & 0.6 & 0.6 & 5 & on\\
  t29 & 70 & 30 & 0.6 & 0.6 & 5 & on\\
  t30 & 1 & 0.5 & 0.6 & 0.6 & 1 & on\\
  t31 & 70 & 30 & 0.6 & 0.6 & 1 & on\\
  t32 & 1 & 0.5 & 0.6 & 0.6 & 0.2 & on\\
  t33 & 70 & 30 & 0.6 & 0.6 & 0.2 & on\\
  t34 & 1 & 0.5 & 0.8 & 0.8 & 5 & on\\
  t35 & 70 & 30 & 0.8 & 0.8 & 5 & on\\
  t36 & 1 & 0.5 & 0.8 & 0.8 & 1 & on\\
  t37 & 70 & 30 & 0.8 & 0.8 & 1 & on\\
  t38 & 1 & 0.5 & 0.8 & 0.8 & 0.2 & on\\
  t39 & 70 & 30 & 0.8 & 0.8 & 0.2 & on\\
  t40 & 1 & 0.5 & 0.2 & 0.2 & 0.04 & on\\
  \enddata
  \tablecomments{The quantities $m_i$ and $R_i$ are the planet masses
    and radii, and $a_{\rm 1,init}$ is the initial semi-major axis of
    planet 1.  At least 5000 simulations were run for each model.}
\end{deluxetable*}

\begin{deluxetable*}{crrrrrrr}
  \tablecaption{Outcome numbers for planet evolution 
simulations.\label{tab:stats}}
  \tablehead{
    \colhead{model name} &
    \colhead{total} &
    \colhead{ejection} &
    \colhead{collision} &
    \colhead{tidal capture} &
    \colhead{two planets} &
    \colhead{stargrazer} &
    \colhead{error\tablenotemark{a}}
  }
  \startdata
  t2 & 5000 & 1450 & 2359 & 0 & 1124 & 6 & 61\\
  t3 & 5000 & 2386 & 1777 & 0 & 830 & 5 & 2\\
  t4 & 5000 & 2436 & 1628 & 0 & 805 & 123 & 8\\
  t5 & 5000 & 1969 & 1377 & 969 & 642 & 33 & 10\\
  t6 & 5000 & 3934 & 373 & 272 & 395 & 25 & 1\\
  t7 & 5000 & 335 & 3804 & 286 & 550 & 25 & 0\\
  t8 & 5000 & 2760 & 1426 & 376 & 303 & 127 & 8\\
  t9 & 5000 & 1234 & 2112 & 1337 & 309 & 1 & 7\\ 
  t10 & 5000 & 4275 & 301 & 212 & 165 & 47 & 0\\
  t11 & 5000 & 401 & 3348 & 1121 & 127 & 0 & 3\\
  t12 & 5000 & 3905 & 612 & 313 & 106 & 64 & 0\\
  t13 & 5000 & 119 & 4653 & 155 & 71 & 1 & 1\\
  t14 & 5000 & 3461 & 1030 & 320 & 41 & 148 & 0\\
  t15 & 5000 & 3213 & 1334 & 302 & 31 & 119 & 1\\
  t16 & 5000 & 834 & 2685 & 1262 & 214 & 1 & 4\\
  t17 & 5000 & 4112 & 428 & 279 & 144 & 37 & 0\\
  t18 & 5000 & 194 & 4042 & 666 & 97 & 0 & 1\\
  t19 & 5000 & 3734 & 771 & 360 & 78 & 57 & 0\\
  t20 & 5000 & 82 & 4809 & 40 & 67 & 2 & 0\\
  t21 & 100000 & 58772 & 36438 & 2389 & 532 & 1856 & 13\\
  t22 & 5000 & 495 & 3204 & 1166 & 133 & 0 & 2\\
  t23 & 5000 & 3957 & 571 & 317 & 118 & 37 & 0\\
  t24 & 5000 & 128 & 4556 & 238 & 76 & 0 & 2\\
  t25 & 5000 & 3517 & 1001 & 365 & 55 & 62 & 0\\
  t26 & 5000 & 43 & 4892 & 4 & 61 & 0 & 0\\
  t27 & 5000 & 2959 & 1743 & 177 & 27 & 94 & 0\\
  t28 & 5000 & 369 & 3548 & 961 & 120 & 0 & 2\\
  t29 & 5000 & 3885 & 657 & 312 & 104 & 41 & 1\\
  t30 & 5000 & 95 & 4715 & 122 & 67 & 0 & 1\\
  t31 & 5000 & 3404 & 1132 & 357 & 47 & 60 & 0\\
  t32 & 5000 & 43 & 4897 & 3 & 57 & 0 & 0\\
  t33 & 5000 & 2828 & 1928 & 134 & 27 & 82 & 1\\
  t34 & 5000 & 265 & 3851 & 788 & 94 & 1 & 1\\
  t35 & 5000 & 3801 & 746 & 325 & 97 & 30 & 1\\
  t36 & 5000 & 97 & 4791 & 49 & 62 & 1 & 0\\
  t37 & 5000 & 3277 & 1256 & 358 & 43 & 63 & 3\\
  t38 & 5000 & 46 & 4898 & 1 & 54 & 1 & 0\\
  t39 & 5000 & 2727 & 2097 & 67 & 27 & 82 & 0\\
  t40 & 5000 & 37 & 4902 & 0 & 60 & 1 & 0\\
\enddata
\tablecomments{We label the outcomes as in
    \citet{2008ApJ...686..621F}.  ``Two planets'' refers to a system
    that has not achieved another outcome (collision, ejection, etc.)
    within the maximum integration time of $5 \times 10^6$ code units
    ($\sim 8\times 10^5$ initial orbital periods of planet 1).}
\tablenotetext{a}{For reference, we give the outcome statistics for each
  model.  The final column includes both errors
  and uncounted outcomes.  A typical error is the system becoming
  dynamically stable (and classified as such) due to integrator energy
  drift.  A typical uncounted outcome is one planet being ejected
  while the other collides with the host star.\vspace{0.5cm}}
\end{deluxetable*}

\section{Detectability of a Planet--Planet Binary}\label{sec:detectability}

\begin{figure}[h]
  \begin{center}
    \includegraphics[width=0.95\columnwidth]{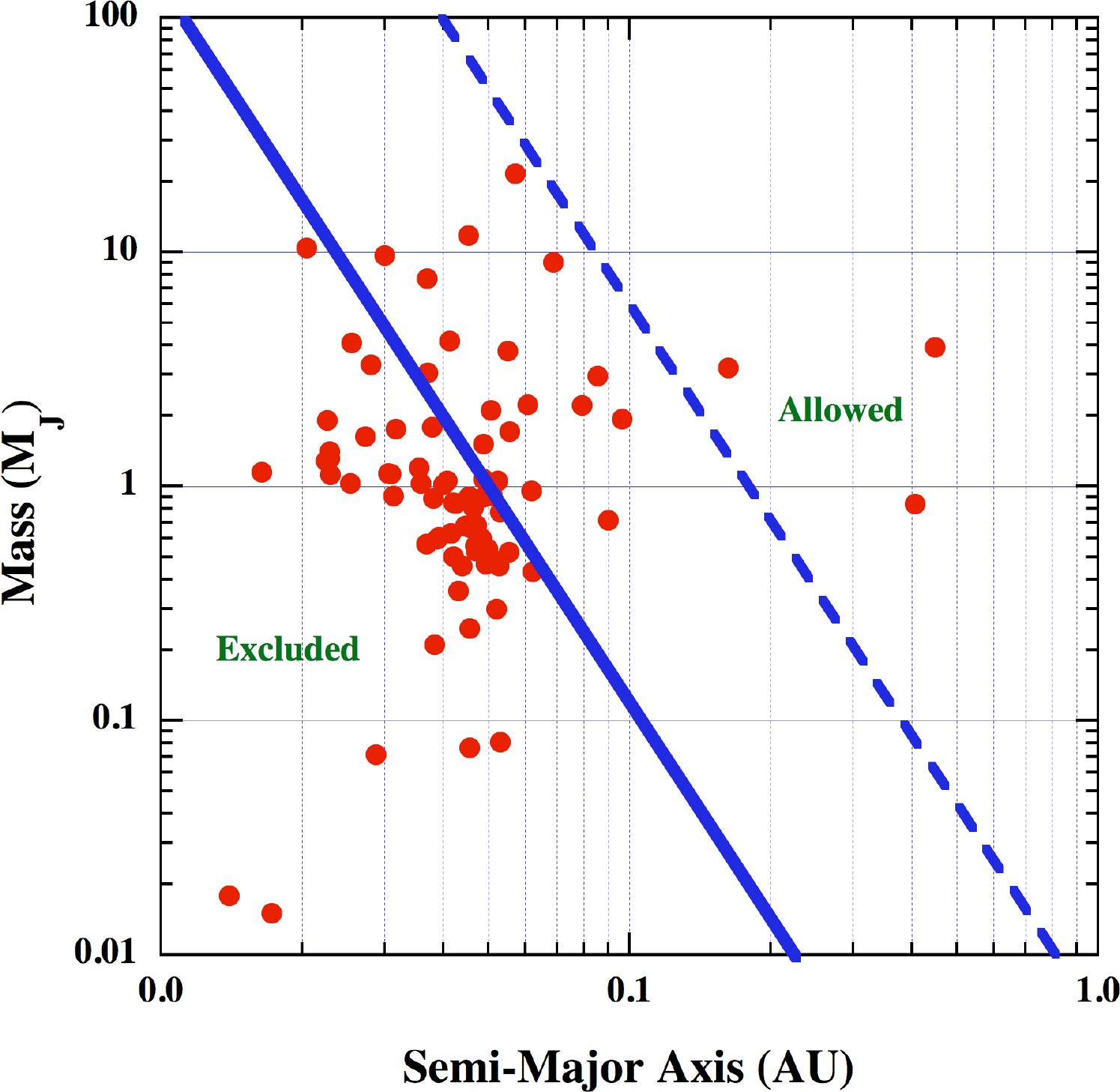}
    \caption{The $\sim 80$ known transiting exoplanet 
systems (taken from ``The Extrasolar Encyclopedia'' 
[http://exoplanet.eu/catalog-all]) 
in the planet mass--semimajor axis plane.  The blue line divides the
      allowed from excluded region of this space if the binary planet
      must fit within 40\% of its Hill sphere and the binary
      components are separated by at least 3 $R_{\rm J}$.  The dashed line
      is slightly more conservative (see text).
      \label{fig:trans}}
  \end{center}
\end{figure}

In this work we have shown that if planets are dynamically active
early in their history, and that our prescription for tidal
interactions is plausibly accurate, then binary planets should form
with comparable frequency with which others are ejected.  Therefore,
since dynamically active planets are the currently favored mechanism
for the formation of eccentric exoplanet orbits
\citep{2008ApJ...686..621F,2008ApJ...686..603J,2008ApJ...686..580C},
we expect that there exist gas-giant binary planets among the
$\sim 420$ that are currently known.  In this section we discuss the
signatures that such hypothetical binary planets would exhibit in
observations of the currently known sample of exoplanets, and
prospects for detecting them in future studies.

Since most of the known exoplanets have been discovered via radial
velocity (hereafter, ``RV'') measurements, the first question to
answer is how large would be the perturbations to the observed Doppler
signature.  Treating the binary planet as an orbital perturbation, 
we derived an expression for the maximum deviations
from a conventional RV curve of the central star 
due to the influence of a binary planet as
\begin{equation}
\Delta V_r \simeq \frac{9}{32} \left(\frac{a}{R_0}\right)^2
\left(\frac{\Omega_0}{\omega}\right)\left(\frac{2m}{M}\right)\,V_0\,,
\end{equation}
where $a$ is the orbital separation of the binary planets, $R_0$ and
$\Omega_0$ are the mean orbital radius of the outer binary and its mean
angular velocity, respectively, $V_0= \Omega_0 R_0$, $\omega$ is the
synodic orbital period of the binary planet, $m$ is the mass of an assumed
equal-mass member of the binary planet, and $M$ is the mass of the
parent star. The mean orbits of the inner and outer binaries have been
taken as circular and coplanar for simplicity.  Since the ratio
$\Omega_0/\omega$ can be expressed in terms of the masses and orbital radii
involved, we can write
\begin{equation}
\Delta V_r \simeq \frac{9}{32} \left(\frac{a}{R_0}\right)^{7/2}
\sqrt{\frac{2m}{M}}\,V_0\,.
\end{equation}
For a binary planet with an orbital separation of $a = 0.4\,R_{\rm
  Hill} \simeq0.4\,(2m/3M)^{1/3}$, i.e., the largest stable
separation, we have
\begin{equation}
\Delta V_r^{\rm max} \simeq \frac{9}{32} \left(\frac{8}{375}\right)^{7/6} 
\left(\frac{2m}{M}\right)^{5/3}\,V_0\,.
\end{equation}
For illustrative values of $m = M_{\rm J}$ and $V_0 = 30$ km s$^{-1}$, the
numerical value of $\Delta V_r^{\rm max}$ is $\sim 0.3$\,cm\,s$^{-1}$, far too
small to be detected in current observations as well as those that are
currently planned for the near future.  On the other hand, for a
binary brown dwarf of, e.g., $50\,M_{\rm J}$, the expected perturbations to
the RV curve would be of order 2 m s$^{-1}$, which would be
detectable.  Thus, it seems fair to say that the existence of
gas-giant planets would not have been noticed in exoplanet RV curves.

We next consider whether binary planets would have been detected among
any of the $\sim 80$ currently known transiting exoplanets.  For these
systems, the transit light curves would be manifestly anomalous and
the presence of two planets would be quite obvious.  In
Fig.\,\ref{fig:trans} we show a plot of 80 currently known transiting
exoplanets in the planet mass--semimajor axis plane.  Note that, as
expected, most of these systems are close to the parent star (i.e.,
$\lesssim 0.1$ AU), thereby enhancing their transit probability.  The
solid blue line indicates where in this plane a binary planet can
fit within 40\% of its Hill sphere and still allow the binary to be
separated by at least 3 times the radius of Jupiter, thereby avoiding
Roche lobe overflow from one planet to another.  In this conservative
set of restrictions, we see that only $\sim$ 1/4 of the systems could
possibly harbor a binary planet.  If we make these constraints only
somewhat more `comfortable' by requiring that the planets be separated
by at least 5 $R_{\rm J}$ and that this separation be less than 1/5 of the
radius of the Hill sphere, the separatrix is shown with a dashed blue
line.  In that case, it seems quite plausible that no more than three
of the currently known transiting systems could contain a binary
planet.  At the moment, the numerical simulations presented in this
work do not place tight constraints on the ratio of highly eccentric
exoplanet orbits and those with binary planets, nor can they predict
an absolute fraction of exoplanets that should be binary.
Nonetheless, it is by no means obvious that any binary planets should
have yet been detected by either RV or transit light curve studies.

It is also possible that binary planets could be detected with
gravitational lensing.  The Einstein radius, $R_{\rm E}$, of a single
planet of mass, $m$, at a distance $D$, and a generic source at
$\sim 2D$, projected back to the lens plane has a physical size
\begin{equation}
R_{\rm E} \simeq \sqrt{\frac{2GMD}{c^2}} \simeq 0.06\,{\rm AU} 
\sqrt{\frac{mD_{\rm kpc}}{M_{\rm J}}} ~.
\end{equation}
Thus, if the planet is actually a binary, it would have to have an
orbital separation comparable to, or greater than $R_{\rm E}$, in
order for its binary nature to be readily revealed in the microlensing
light curves.  For comparison, the maximum separation between binary
components of a Jupiter mass at an AU from a 1 $M_\odot$ star is
$\sim 0.03\,$AU.

Probably the best hope for detecting binary planets is to obtain a
larger sample of transiting exoplanets, yielding a substantial number
(e.g., $\gtrsim$ a dozen) at distances from the parent star of
$\gtrsim0.4$\,AU.  The Kepler mission should ultimately provide such a
sample.  At these larger distances for transiting exoplanets, there is
more phase space to fit stable binaries within their Hill sphere and
to allow larger separations between the binary components.  

If gas-giant binary planets are ultimately discovered, they would
represent a strong corroboration of the dynamically active scenario,
since gas giant binaries formed via tidal capture are a natural
outcome of dynamically active systems.  On the other hand, if no
binary planets are detected, even with a larger sample of transiting
exoplanets, it may simply be that our treatment of the tidal capture
process is too optimistic.  Parameters of the tidal capture process
could be constrained by comparing $N$-body simulations that allow for
tidal capture with observed planet eccentricity distributions, with
the restriction that no observable planet--planet binaries be formed.


\acknowledgements

JMF acknowledges support from Chandra
Postdoctoral Fellowship Award PF7-80047.

\begin{appendix}
\section{Use of the Fewbody Dynamical Code}

In our numerical simulations we use the
\Fewbody\ integrator, which is designed for strong small-$N$-body
gravitational encounters \citep{2004MNRAS.352....1F}.
To test \Fewbody's suitability for dynamically active planetary
systems, we compare our scattering calculations with the work of
\citet{2008ApJ...686..621F}, who studied planet--planet scattering
with the aim of explaining the high eccentricities of some observed
extrasolar planetary systems.  As with most studies of planet--planet
scattering, they used a mixed variable symplectic method modified to
treat close encounters
\citep{1991AJ....102.1528W,1999MNRAS.304..793C}.  When this method is
applied to a two-planet system, a close encounter between the planets
results in {\em all} orbital motion being integrated with a standard,
non-symplectic integrator (e.g., Bulirsch-Stoer).  Since the
two-planet systems we study are dynamically active and hence quickly
result in close approaches, their evolution should be faithfully
treated (in a statistical sense) with the adaptive, but
non-symplectic, integration algorithm in \Fewbody.

\begin{figure}
\begin{center}
    \includegraphics[width=0.45\columnwidth]{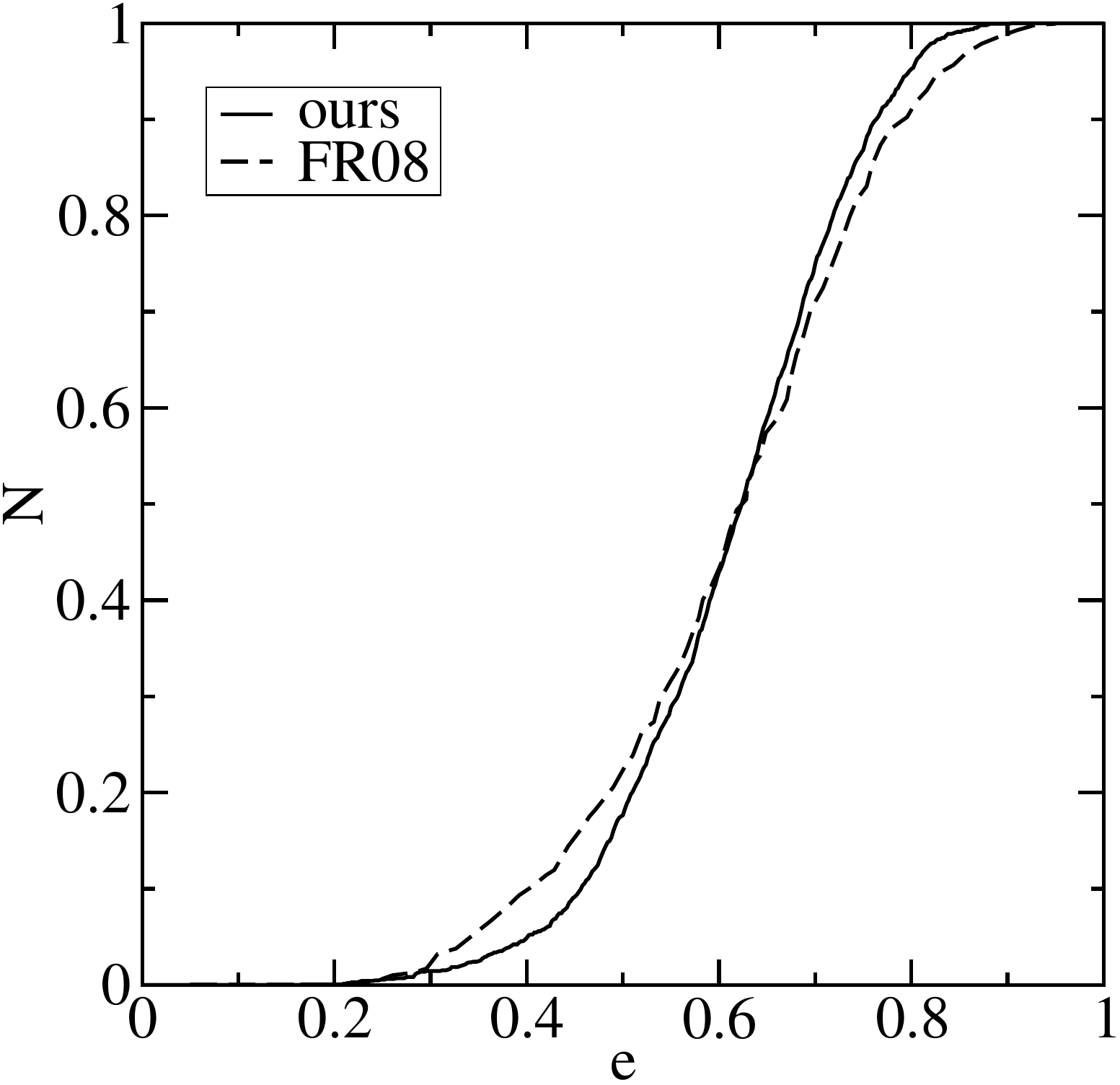} 
\hglue0.3cm
   \includegraphics[width=0.45\columnwidth]{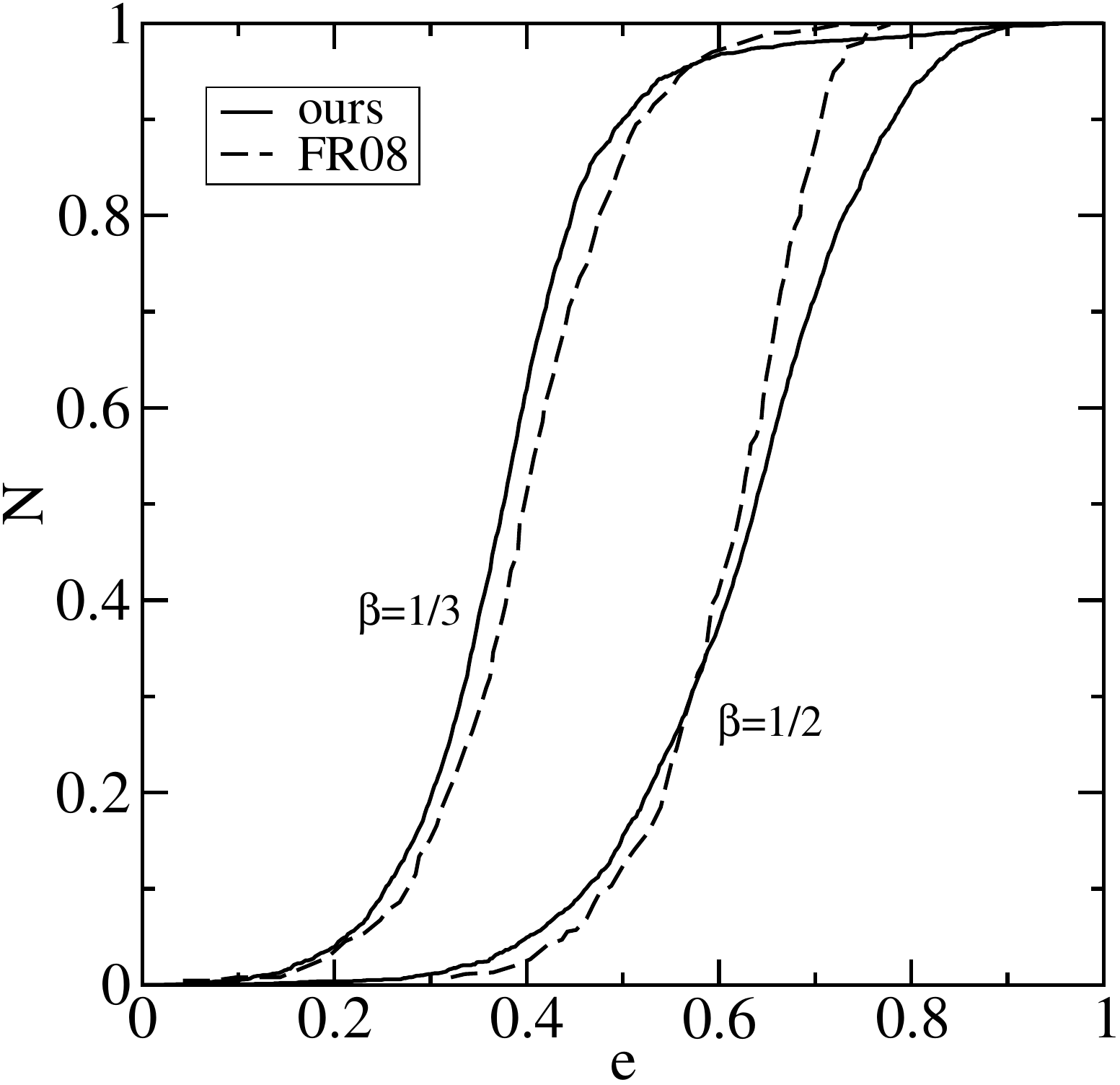}
  \caption{{\em Left Panel:} Comparison with Fig.\ 2 of
    \citet{2008ApJ...686..621F}, the cumulative eccentricity
    distribution of the remaining planet after a planet is ejected for
    the case $m_1/M_\star=m_2/M_\star=10^{-3}$.  As in
    \citet{2008ApJ...686..621F}, we set the stellar mass to $M_\star=1
    \, M_\sun$ and the planet masses to $m_i=10^{-3} \, M_\sun$.
    {\em Right Panel:} Comparison with Fig.\ 3 of
      \citet{2008ApJ...686..621F}, the cumulative eccentricity
      distribution of the remaining planet after a planet is ejected
      for $\beta\equiv m_1/m_2=1/2$ and $\beta=1/3$.  As in
      \citet{2008ApJ...686..621F}, we set $m_1+m_2=6\times 10^{-3}
      M_\star$ and $M_\star=1 \, M_\sun$.
      \label{fig:fr08fig2}}
\end{center}
\end{figure}

Fig.\ \ref{fig:fr08fig2} (left panel) shows a comparison of our
numerical method with Fig.\ 2 of \citet{2008ApJ...686..621F}, who
integrated the evolution of dynamically active two-planet systems.
Specifically, the cumulative eccentricity distribution of the
remaining planet after a planet is ejected is shown for the case
$m_1/M_\star=m_2/M_\star=10^{-3}$, where $m_i$ is the mass of planet
$i$ and $M_\star$ is the mass of the host star.  By convention, planet
1 initially orbits with semi-major axis $a_{\rm 1,init} < a_{\rm
  2,init}$.  Our initial conditions (including planet eccentricities,
relative inclinations, and randomization of various orbital elements)
are the same as in \citet{2008ApJ...686..621F}.  Most notably for this
case, we similarly set the stellar mass to $M_\star=1 \, M_\sun$ and
the planet masses to $m_i=10^{-3} \, M_\sun$.  As is clear from the
figure, the eccentricity distributions are fairly similar -- in fact,
the median eccentricity is the same in both distributions.  The
differences are most apparent at low and high eccentricities.  Part of
the discrepancy may arise from our differing definitions of ejection.
In \citet{2008ApJ...686..621F}, a planet straying beyond $2000 a_{\rm
  1,init}$ is declared to have been ejected, while in \Fewbody\ a
planet is not considered ejected until it is formally unbound from and
receding from the remaining masses, and is sufficiently weakly tidally
coupled to the remaining masses that it could never become bound
again.  Another source for the discrepancy may be the accuracy of the
mixed variable symplectic method for close star--planet approaches
used in \citet{2008ApJ...686..621F}.  As pointed out in
\citet{1999AJ....117.1087R}, the method introduces artificial chaos
for close star--planet approaches.  Finally, a remaining source for
the discrepancy may be the calculation stopping time.  We used a
stopping time of $5 \times 10^6$ code units ($\sim 8\times 10^5$
initial orbital periods of planet 1) for all calculations.
\citet{2008ApJ...686..621F} used a stopping time between $5 \times
10^6$ and $2 \times 10^7$ that is an unspecified function of the
planet masses.

Fig.\ \ref{fig:fr08fig2} (right panel) shows a comparison of our
results with Fig.\ 3 of \citet{2008ApJ...686..621F}.  Shown is the
cumulative eccentricity distribution of the remaining planet after a
planet is ejected for $\beta\equiv m_1/m_2=1/2$ and $\beta=1/3$.  As
in \citet{2008ApJ...686..621F}, we set $M_\star=1 \, M_\sun$ and
$m_1+m_2=6\times 10^{-3} M_\star$ to speed up the evolution.  (As
shown in \citet{2008ApJ...686..621F}, the final eccentricity
distribution is not sensitive to the ratio $(m_1+m_2)/M_\star$ for
$m_i/M_\star \sim 10^{-3}$, but the time to ejection decreases as
$(m_1+m_2)/M_\star$ increases.)  While the agreement between the
eccentricity distributions is not perfect, our results agree fairly
well with those of \citet{2008ApJ...686..621F}, and reproduce the
dependence of the eccentricity distribution on $\beta$.  Additionally,
eccentricities above 0.8 are very rare, as found in
\citet{2008ApJ...686..621F}.


\end{appendix}


\bibliographystyle{apj}
\bibliography{apj-jour,main,saul,philipp}

\clearpage

\end{document}